\documentclass[final]{acmtrans2e}

\usepackage{graphicx}
\usepackage{ulem}
\usepackage{color}
\usepackage{latexsym}
\usepackage{amsmath}
\usepackage{amsfonts}

\newtheorem{theorem}{Theorem}[section]

\newtheorem{corollary}[theorem]{Corollary}
\newtheorem{proposition}[theorem]{Proposition}
\newtheorem{lemma}[theorem]{Lemma}
\newdef{definition}[theorem]{Definition}
\newdef{remark}[theorem]{Remark}

\newtheorem{exa}[theorem]{Example}
\newenvironment{example}{\begin{exa}}{\end{exa}}

\newtheorem{exe}{Exercise}

\newcommand{\la}{\langle}
\newcommand{\ra}{\rangle}

\newcommand{\rrarrow}{\longrightarrow}

\long\def\comment#1{}

\title{Unfolding in CHR}

\author{Maurizio Gabbrielli\\
  Universit\`{a} di Bologna
  \and
  Maria Chiara Meo\\
  Universit\`a ``G. D'Annunzio'' di Chieti-Pescara
  \and
  Paolo Tacchella\\
   Universit\`{a} di Bologna}

\begin{abstract}
Program transformation is an appealing technique which allows to
improve  run-time efficiency, space-consumption and more generally
to optimize a given program. Essentially it consists of a sequence
of syntactic program manipulations which preserves some kind of
semantic equivalence.  One of the basic operations which is used
by most program transformation systems is unfolding  which
consists in the replacement of a procedure call by its definition.
While there is a large body of literature on transformation and
unfolding of sequential programs, very few papers have addressed
this issue for concurrent languages and, to the best of our
knowledge, no other has considered unfolding of CHR programs.

This paper defines a correct unfolding system
for CHR programs. We define an unfolding rule, show its
correctness and discuss some conditions which can be used to
delete an unfolded rule while preserving the program meaning.
We prove that confluence and termination properties
are preserved by the above transformations.
\end{abstract}

\begin{document}

            \begin{bottomstuff}
             Author's address:
             \newline
             Maurizio Gabbrielli,
  Dipartimento di Scienze dell'Informazione, Mura A. Zamboni 7, 40127 Bologna, Italy.
  {\tt gabbri@cs.unibo.it.}
  \newline Maria Chiara Meo, Dipartimento di Scienze, Viale Pindaro 42,
  65127 Pescara, Italy.
{\tt cmeo@unich.it.}
 \newline Paolo Tacchella, Dipartimento di Scienze dell'Informazione, Mura A. Zamboni 7, 40127 Bologna, Italy.
{\tt Paolo.Tacchella@cs.unibo.it} \end{bottomstuff}

\maketitle
\section{Introduction}
Program transformation was initially developed as a
technique which assist in writing correct and efficient programs \cite{BD77}.
Said technique consists of many intermediate transformation steps until the final
one is reached. Every transformed program is equivalent (gives the same results) of
the initial one, when an input is fixed. The transformation between various algorithms
which compute Fibonacci succession can be considered as an example of program transformation.
In fact, the time complexity of the previous succession ranges from the exponential to the logarithmic
depending on the chosen  algorithm \cite{SP95}.

CHR is a general purpose \cite{SSD05d}, declarative, concurrent, committed-choice
constraint logic programming language, consisting of guarded rules, which
transform multisets of atomic formulas (constraints) into simpler ones to the point of
exhaustion \cite{Fru06}, that was initially designed for writing
constraint solvers \cite{Fru98,FA03}.
There is nowadays a very  large literature on
CHR, ranging from theoretical aspects to implementations and
applications. \\
In fact, the website
http:/\!/www.cs.kuleuven.ac.be/\,\lower 3.5pt\hbox{\~{}} \,dtai/projects/CHR/
reports more than 1000 papers mentioning CHR. However,
only a few papers, notably \cite{FH03,Fru04,SSD05b}, consider  source
to source transformation of CHR programs. This is not surprising,
since program transformation is in general very difficult for
(logic) concurrent languages and in case of CHR it is even more
complicated, as we discuss later.

While  \cite{Fru04}  focuses on specialization of a program for a
given  goal, here we consider unfolding.  This is a basic
operation of any source to source transformation (and
specialization) system and essentially consists in the replacement
of a procedure call by its definition. While this operation can be
performed rather easily for sequential languages, and indeed  in
the field of logic programming it was first investigated by Tamaki
and Sato more than twenty years ago \cite{TS84}, when considering
logic concurrent languages it becomes quite difficult to define
reasonable conditions which ensure its correctness. This is mainly
due to three problems. The first one is the presence of guards in the rules.
Intuitively, when
unfolding a rule $r$ by using a rule $v$ (i.e. when replacing in
the body of $r$ a ``call'' of a procedure by its definition $v$)
it could happen that some guard in $v$ is not satisfied
``statically'' (i.e. when we perform the unfold), even though it
could become satisfied later when the unfolded rule is actually
used. If we move the guard of $v$ in the unfolded version of $r$
we can then loose some computations (because the guard is
anticipated). This means that if we want to preserve the meaning
of a program we cannot replace the rule $r$ by its unfolded
version, and we have to keep both the rules.
The second source of difficulties consists in matching substitution
mechanism. Only the variables in the atoms of the head of a rule
$r$ can be instantiated to become equal to the goal terms following
the previous mechanism. From the other side, the unification mechanism
permits also the instantiation of the variables in the atoms of the
goal. Considering the matching substitution, the deletion of $r$, when
a rule $v$ could be used to unfold $r$ if strong enough hypotheses
would be considered, can cause computation loss also if $r$ is unfolded
by another rule $v'$. Finally, for CHR, the situation is further complicated by
the presence of multiple heads in the rules.
In fact, let $B$ be the body of a rule $r$ and let $H$ be the (multiple)
head of a rule $v$, which can be used to unfold $r$, we cannot be sure
that at run-time all the atoms in $H$ will be used to rewrite $B$,
since in general $B$ could be in a conjunction with other atoms even
though the guards are satisfied. This technical point, that one can legitimately find
obscure now, will be further clarified in Chapter
\ref{sec:safty-rule-deletion}.

Despite these technical problems, the study of unfolding techniques
for concurrent languages, and for CHR in particular, is important
as it could lead to significant improvements in the efficiency and
in non-termination analysis of programs.

In this paper we then define an unfolding rule for CHR programs
and show that it preserves the semantics of the program in terms
of qualified answers, a notion already defined in the literature
\cite{Fru98}.
We also provide a syntactic condition which allows to replace in a
programs a rule by its unfolded version  while preserving
qualified answers. Even though the idea of the unfolding is
straightforward, its technical development is complicated by the
presence of guards, multiple heads and matching substitution,
as previously mentioned. In
particular, it is not immediate to identify conditions which
allow to replace the original rule by its unfolded version.
Moreover, a further reason of complication comes from the fact
that we consider the reference semantics (called $\omega_t$)
defined in \cite{DSGH04} which avoids trivial non termination by
using a, so called, token store or history. Due
to the presence of this token store,  in order to define correctly
the unfolding we have to slightly modify the syntax of CHR programs
by adding to each rule  a local token store. The resulting programs
are called annotated and we define their semantics by providing a
(slightly) modified version of the semantics $\omega_t$, which is
proven to preserve the qualified answers. Finally, the maintenance
of confluence and termination of property between the original and
the ones, which are modified following the above techniques, is proven.

The remaining of this paper is organized as follows. Next section
contains some notations used in the paper and the syntax of CHR.
The operational semantics of $\omega_t$ \cite{DSGH04} and of the
modified semantics $\omega'_t$  are presented  in
Section~\ref{sec:semantics}. Section~\ref{sec:unfolding} defines
the unfolding rule and prove its correctness.
Section~\ref{sec:safty-rule-deletion}  discuss the problems
related to the replacement of a rule by its unfolded version and
gives a correctness condition which holds for a specific class of
rules. Then Section~\ref{sec:confluence&termination} proves that
confluence and termination are preserved by the program modifications
introduced. Finally Section~\ref{sec:conclusion_and_future} concludes
by discussing also some related work.

\section{Preliminaries}\label{sec:notation}

In this section we introduce the syntax of CHR and some notations
and definitions we will need in the paper. CHR uses two kinds
of constraints: the built-in and the CHR ones, also called
user-defined.

Built-in constraints are defined by
$$ c ::= d \ | \ c \wedge c \ | \ \exists_x c $$
where $d$ is an atomic built-in constraint\footnote{We could
consider more generally first order formulas as built-in
constraints, as far as the results presented here are concerned.}.
These constraints are handled by an existing solver and we assume
given a (first order) theory CT which describes their meaning.
We assume also that built-in constraints contain $=$ which is described, as usual, by the Clark Equality Theory.

We use $c,d$ to denote built-in constraints, $h,k,s, p, q$ to denote CHR
constraints and $a,b, g, f$ to denote both built-in and user-defined
constraints (we will call these generically constraints).  We also
denote by ${\tt false}$ any inconsistent (conjunction of)
constraints and by ${\tt true}$ the empty set of
constraints. The capital versions will be used to denote multisets
(or sequences) of constraints.

The notation $\exists_{-V}\phi$, where $V$ is a set of variables,
denotes the existential closure of a formula $\phi$ with the
exception of the variables in $V$ which remain unquantified.
$Fv(\phi)$ denotes the free variables appearing in $\phi$.
Moreover, if $\bar t= t_1, \ldots t_m$ and $\bar t'= t'_1, \ldots t'_m$ are sequences of terms then the notation
$p(\bar t) = p'(\bar t')$
represents the set of equalities $\,t_1=t'_1, \ldots, t_m=t'_m\,$ if $p=p'$, and it is undefined otherwise.
Analogously, if $H=h_1, \ldots, h_k$ and $H'= h'_1, \ldots, h'_k$ are sequences of constraints, the notation
$H=H'$ represents the set of equalities $h_1=h'_1, \ldots, h_k=h'_k$.
Finally, multiset union is represented by symbol $\uplus$.

\subsection{CHR syntax}\label{sec:syntax}
As shown by the following definition, a  \emph{CHR program}
consists of a set of rules which can be divided into three types:
\emph{simplification}, \emph{propagation} and \emph{simpagation}
rules. The first kind of rules is used to rewrite
CHR constraints into simpler ones, while second one allows to add new redundant
constraints which may cause further simplification. Simpagation rules allow to represent both simplification and propagation rules.

\begin{definition}\textsc{CHR Syntax} \cite{Fru98}.
A CHR program is a finite set of CHR rules.
There are three kinds of CHR rules:

\noindent{A \textbf{simplification} rule has the form: $${\it r}@ H
\Leftrightarrow C  \,|\, B$$}\\
\noindent{A \textbf{propagation} rule has the form:
$${\it r}@ H \Rightarrow C  \,|\,  B$$}\\
\noindent{A \textbf{simpagation} rule has the form:
$${\it r}@ H_1
\setminus H_2 \Leftrightarrow C  \,|\,  B,$$}\noindent where ${\it r
}$ is a unique identifier of the rule, $H$, $H_1$ and $H_2$ are
sequences of user-defined constraints (called heads), $C$ is a
possibly empty multiset of built-in constraints (guard) and $B$ is
a possibly empty multiset of (built-in and user-defined)
constraints (body). A \emph{CHR goal} is a multiset of (both
user-defined and built-in) constraints.
\end{definition}

A  \emph{simpagation} rule can simulate both simplification and
propagation rule by considering, respectively, either $H_1$ or
$H_2$ empty (with $(H_1,H_2)\neq \emptyset$). In the following we
will then consider in the formal treatment only simpagation rules.

When considering unfolding we need to consider a slightly
different syntax,  where rule identifiers are not necessarily
unique, atoms in the body are associated with an identifier,
that is unique in the rule, and
where each rule is associated  with a local token store $T$.
More precisely, we define an identified CHR constraint  (or
identified atom) $h\#i$ as a CHR constraint $h$, associated with
an integer $i$ which allows to distinguish different copies of the
same constraint.

\begin{definition}\textsc{CHR Annotated syntax.}
Let us define a token as an object of the form $r@i_1, \ldots, i_l$, where $r$
is the name of a rule and $i_1, \ldots, i_l$ is a sequence of identifiers.
A token store (or history) is a set of tokens.

\noindent An \textbf{annotated} rule has then the form:
$${\it r}@ H_1
\setminus H_2 \Leftrightarrow C  \,|\,  \tilde B;T$$ where ${\it r }$
is an identifier, $H_1$ and $H_2$ are
sequences of user-defined constraints, $\tilde B$
is a sequence of built-in and identified CHR constraints
such that different (occurrences of) CHR constraints have different identifiers,
and $T$ is a token store, called the local token store of rule $r$.
An annotated CHR program is a finite set of annotated CHR rules.
\end{definition}
We will also use the functions {\em chr(h$\#$i)=h} and the overloaded
function {\em id(h$\#$i)=i}, [and $id(r@i_1,\ldots,i_l)=\{i_1, \ldots, i_l\}$]
possibly extended to sets and sequences of identified CHR constraints [or tokens] in
the obvious way. Given a goal $G$, we denote by $\tilde G$ one of the
possible identified versions of $G$. $Goals$ is the set of all (possibly
identified) goals.

Intuitively, identifiers are used to distinguish different
occurrences of the same atom in a rule. The identified atoms can
be obtained by using a suitable function which associates a
(unique) integer to each atom. More precisely, let $B$ be a goal
which contains $m$ CHR-constraints. We assume that the function
$I_n ^{n+m}(B)$ identifies each CHR constraint in $B$ by
associating to it a unique integer in $[n+1,m+n]$ according to the
lexicographic order.

On the other hand, the token store allows to memorize  some tokens,
where each token describes which (propagation) rule has been used
for reducing which identified atoms.
As we discuss
in the next section, the use of this information was originally proposed
in \cite{Abd97} and then further elaborated in the semantics defined
in \cite{DSGH04} in order to avoid trivial  non termination
arising from the repeated application of the same propagation rule
to the same constraints. Here we simply incorporate this
information in the syntax, since we will need to manipulate it in
our unfolding rule.

Given a CHR program $P$, by using the function $I_n ^{n+m}(B)$ and
an initially empty local token store we can construct its annotated
version as the next definition explains.

\begin{definition}
Let $P$ be a CHR program. Then its annotated version is defined as follows:
\[\begin{array}{lll}
  Ann(P)=\{&{\it r}@ H_1 \setminus H_2 \Leftrightarrow C  \,|\,  I_0
^{m}(B);\emptyset \mid& \\
& {\it r}@ H_1 \setminus H_2 \Leftrightarrow C
 \,|\,
B \in P \mbox{ and } &\\
&\mbox{$m$ is the number of CHR-constraints in $B$}&\}.
\end{array}
\]
\end{definition}

\noindent {\bf Notation} \\
In the following examples, given a
(possibly annotated) rule
$${\it r}@ H_1 \setminus
H_2 \Leftrightarrow C  \,|\,  B(;T),$$   we write it as
$${\it r}@ H_2 \Leftrightarrow C  \,|\,  B(;T),$$ if $H_1$ is empty and
we write it as
$${\it r}@ H_1 \Rightarrow C  \,|\,  B(;T),$$ if $H_2$ is empty.

That is, we maintain also the notation previously introduced for
simplification and propagation rules. Moreover, if $C=
\texttt{true}$, then $\texttt{true}  \,|\, $ is omitted.
Finally,  if in an annotated rule the token store is empty we
simply  omit it.

\section{CHR operational semantics}\label{sec:semantics}

This section introduces the reference semantics $\omega_t$
\cite{DSGH04}, in particular the variant that modifies the token
set only after the application of a propagation rule (for the sake
of simplicity, we omit indexing the relation with the name of the
program).

Afterward we define a slightly different operational  semantics,
called $\omega_t'$, which considers annotated programs and which
will be used to prove the correctness of our unfolding rules (via some form of
equivalence between  $\omega_t'$ and $\omega_t$).

\begin{table*}[tbp]
\caption{The transition  system $T_{\omega_t}$ for the $\omega_t$ semantics}
\centering
\label{omega-t}

\begin{tabular}{lll}
\hline\noalign{\smallskip}
&\mbox{   }&\mbox{   }
\\
\mbox{\bf Solve} &  $\displaystyle{\frac {CT \models
c \wedge C \leftrightarrow C' \hbox{ and c is a built-in
constraint}}  {\la \{c\}\uplus G, \tilde S,C, T \ra_n
\rrarrow_{\omega_t} \la G, \tilde S,C', T \ra_n}} $

&\mbox{ }
\\
&\mbox{   }&\mbox{   }
\\

\mbox{\bf Introduce}& $\displaystyle{\frac { \hbox{h
is a user-defined constraint}}{ \la \{h \}\uplus G, \tilde S,C, T
\ra_n \rrarrow_{\omega_t} \la G,\{ h\#n \} \cup \tilde S ,C, T
\ra_{n+1} }}$ &\mbox{ }
\\
&\mbox{   }&\mbox{   }
\\

\mbox{\bf Apply}& $\displaystyle{\frac {
\begin{array}{c}
{\it r}@H'_1 \setminus H'_2 \Leftrightarrow D  \,|\,  B \in P \ \ \
x = Fv(H'_1, H'_2)  \\
CT \models C \rightarrow \exists _{x}
((chr(\tilde H_1, \tilde H_2)=(H'_1,  H'_2))\wedge D)
\end{array}
}
{\displaystyle
\begin{array}{c}
\la G, \{\tilde H_1\} \cup \{ \tilde H_2\}\cup
\tilde S,C, T \ra_n \rrarrow_{\omega_t}\\ \la B\uplus G,\{ \tilde
H_1\} \cup \tilde S, (chr(\tilde H_1, \tilde H_2)=(H'_1,
H'_2))\wedge C, T' \ra_n
\end{array} }}$&\mbox{ }
\\
&\mbox{   }&\mbox{   }
\\
&\mbox{where }${\it r}@id (\tilde  H_1,  \tilde H_2) \not \in T $
\mbox{ and }\\
&$T'=T \cup \{{\it r}@id ( \tilde H_1,  \tilde H_2)\}
\mbox{ if } \tilde H_2=\emptyset \mbox{ otherwise } T'=T. $&\mbox{
}
\\
&\mbox{   }&\mbox{   }
\\
\noalign{\smallskip}\hline
\end{tabular}
\end{table*}

We describe the operational semantics $\omega_t$, introduced in
\cite{DSGH04}, by using a transition system
\[T_{\omega_t}= ({\it Conf_t},
\rrarrow_{\omega_t}).\] Configurations in ${\it Conf_t}$ are
tuples of the form $\langle G,\tilde  S,c, T\rangle_n$ with the
following meaning. The \emph{goal} $G$ is a multiset of constraints to
be evaluated. The \emph{CHR constraint store} $\tilde S$ is the set of
identified CHR constraints that can be matched with the head of the
rules in the program $P$. The \emph{built-in constraint store} $c$ is a
conjunction of built-in constraints. The {\em propagation history}
$T$ is a set of tokens of the form $r@i_1, \ldots, i_l$, where $r$
is the name of the applied propagation rule and $i_1, \ldots, i_l$ is the
sequence of identifiers associated to the constraints to which the
head of the rule is applied. This is needed to prevent trivial
non-termination for propagation rules. If one do not consider
tokens (as in the original semantics of \cite{Fru98}) it is clear
from the transition system that if a propagation rule can be
applied once it can be applied infinitely many times thus
originating an infinite computation (no fairness assumptions are
made here). On the other hand, by using tokens one can ensure that
a propagation rule is used to reduce a sequence of constraints
only if the same rule has not been used before on the same
sequence of constraints, thus avoiding trivial infinite
computations (arising from the application of the same rule to the
same constraints). As previously mentioned, the first idea of using
a token store to avoid trivial non termination was described in
\cite{Abd97}. Finally the \emph{counter} $n$ represents the next
free integer which can be used to number a CHR constraint.

Given a goal $G$, the  {\em initial configuration} has the form
\[\langle G,\emptyset,{\tt true}, \emptyset \rangle_1.\]
A {\em final configuration} has either the form $\langle  G',
\tilde S, {\tt false} , T\rangle_n$ when it is {\em failed} or it
has the form $\langle \emptyset,\tilde S, c,T \rangle_n$ when it
represents a successful termination (since there are no more
applicable rules).

The relation $\rrarrow_{\omega_t}$ (of the transition system
of the operational semantics $\omega_t$) is defined  by the rules
in Table \ref{omega-t}: the \textbf{Solve} rule moves a
built-in constraint from goal store to the built-in constraint
store; the \textbf{Introduce} identifies and moves a CHR (or used
defined) constraint from the goal store to the CHR constraint
store and the \textbf{Apply} rule chooses a program rule $r$, for
which matching between constraints in CHR store and the ones in
the head of $r$ exists, it checks that the guard of $r$ is
entailed by the built-in constraint store, considering the
matching substitution, and it verifies that the token that would be
eventually added by \textbf{Apply} in the token store is not already
present, than it fires the rule. After the application of $r$
the constraints which match with the right hand side of the head
of $r$ are deleted from $\tilde S$, the body of $r$ is added to
the CHR constraint store and the matching substitution between
the head of $r$ and the atoms in $\tilde S$ is added to the
built-in constraint store.

\subsection{The  modified semantics $\omega_t'$}
We now define the semantics $\omega_t'$ which considers annotated rules.
This semantics differs from $\omega_t$ in two aspects.

First, in $\omega_t'$
the goal store and the CHR store are fused in a unique generic \emph{store}, where
CHR constraints are immediately labeled. As a consequence, we do not
need anymore the Introduce rule and every CHR
constraint in the body of an applied rule is immediately
utilizable for rewriting.

The second difference concerns the shape of the rules. In fact,
each annotated rule $r$ has  a local token store (which can be
empty) that is associated to it and which is used to keep trace of
the propagation rules that are used to unfold the body of $r$.
Note also that here, differently from the case of the propagation
history in $\omega_t$, the token store associated to
 the real computation can be updated
by adding more tokens at once (because an unfolded rule with many
token in its local token store has been used).

In order to define formally $\omega_t'$
we need a function $inst$  which updates
the formal identifiers of a rule to the actual computation ones and it
is defined as follows.

\begin{definition}\label{definst}
 Let $Token$ be the set of all possible token
set and let $\mathbb{N}$ be the set of natural numbers. We denote
by $inst: Goals \times \{Token\}\times \mathbb{N}  \rightarrow
Goals \times \{Token\}\times \mathbb{N}$ the function such that
$inst(\tilde B,T,n)=(\tilde B', T',m$), where
\begin{itemize}
    \item $\tilde B$ is an identified CHR goal,
    \item $(\tilde B', T') $ is obtained from $(\tilde B, T)$ by
    incrementing each identifier in $(\tilde B, T)$ with $n$ and
    \item $m$ is the greatest identifier in $(\tilde B', T')$.
\end{itemize}
\end{definition}

We describe now the operational semantics $\omega_t'$  for annotated CHR
programs  by using, as usual, a transition system
\[T_{\omega'_t}= ({\it Conf'_{t}}, \rrarrow_{\omega'_t}).\] Configurations in
${\it Conf'_t}$ are tuples of the form $\langle \tilde S,c,
T\rangle_n$ with the following meaning.  $\tilde S$ is the set of
identified CHR constraints that can be matched with rules in the
program $P$ and built-in constraints. The built-in constraint
store $c$ is a conjunction of built-in constraints and $T$ is a
set of tokens, while the counter $n$ represents the last integer
which  was used to number  the CHR constraints in $\tilde S$.

Given a goal $G$, the  {\em initial configuration} has the form
\[\langle I_0^m(G),{\tt true}, \emptyset \rangle_m,\]
where $m$ is the number of CHR constraints in $G$. A {\em final
configuration} has either the form $\langle \tilde S, {\tt false}
, T\rangle_n$ when it is {\em failed} or it has the form $\langle
\tilde S, c,T \rangle_n$ when it represents a successful
termination, since there are no more applicable rules.

The relation $\rrarrow_{\omega'_t}$ (of the transition system of
the operational semantics $\omega'_t$) is defined  by the rules in
Table \ref{tab:operational-semantics}.  Let us discuss briefly
the rules.

\begin{description}
\item[Solve']{moves a built-in constraint from the store to the
built-in constraint store;}

\item[Apply']{uses the rule $r@H_1'\backslash H_2'\Leftrightarrow
D \, |\, \tilde B; T_r$ provided that exists a matching
substitution $\theta$ such that $chr(\tilde H_1,\tilde H_2)=
(H_1',H_2')\theta$, $D$ is entailed by the built-in constraint
store of the computation and $r@id(\tilde H_1, \tilde H_2)\not \in
T$; $\tilde H_2$ is replaced by $\tilde B$, where the identifier
are suitably incremented by $inst$ function and $chr (\tilde
H_1,\tilde H_2)= (H_1',H_2')$ is added to built-in constraint
store.}
\end{description}

\begin{table*}[t]
\caption{The transition system $T_{\omega'_t}$ for the $\omega'_t$ semantics}
\centering
\label{tab:operational-semantics}
$$
\begin{array}{llll}

\hline\noalign{\smallskip}
&&&\\

\mbox{   }&\textbf{Solve'}&\displaystyle\frac{CT\models c\wedge
C\leftrightarrow C' \mbox{ and } c \mbox{ is a built-in
constraint}} {\langle\{c\}\cup \tilde G, C,
T\rangle_n\rrarrow_{\omega'_t}
\langle \tilde G, C', T\rangle_n}&\mbox{   }\\

&&&\\

&\textbf{Apply'}&\displaystyle\frac{
\begin{array}{c}
r@H'_1\backslash H'_2
\Leftrightarrow D\, | \, \tilde B ; T_r\in P,  \quad x=Fv(H'_1,
H'_2)\quad \\
CT\models C\rightarrow \exists_x((chr(\tilde H_1,
\tilde H_2)=(H'_1, H'_2))\wedge D)
\end{array} } {
\begin{array}{c}
\langle \tilde H_1\cup \tilde
H_2\cup \tilde G, C, T \rangle_n\rrarrow _{\omega'_t} \\
\langle \tilde B' \cup \tilde H_1\cup \tilde G, (chr(\tilde H_1,
\tilde H_2)=(H'_1, H'_2)\wedge C, T'\rangle_{m}
\end{array}
}&\mbox{   }\\

&&&\\

&&\mbox{where  } (\tilde B', T'_r,m )= inst(\tilde B,T_r,n);
r @id(\tilde H_1, \tilde H_2)\not\in T \mbox{ and } &\mbox{   }\\
&&T'=T\cup \{r @id(\tilde H_1, \tilde H_2)\} \cup T'_r \mbox{ if }
\tilde H_2=\emptyset \mbox{ otherwise } T'=T\cup T'_r
.&\mbox{   }\\
&&&\\
\noalign{\smallskip}\hline
\end{array}
$$
\end{table*}

In order to show the equivalence of the semantics $\omega_t$ and
$\omega'_t$ we now define the  notion of observables that we
consider: these are the ``qualified answers'' (already used in
\cite{Fru98}).

\begin{definition}\textsc{(Qualified answers)}. Let $P$ be a CHR program and let
$G$ be a goal. The set $\mathcal{QA}_P(G)$ of
qualified answers for the query $G$ in the program $P$ is defined
as follows:
$$
\hspace*{-0.2cm}\begin{array}{l}
\mathcal{QA}_P(G) =\\
\hspace{0.3cm} \{\exists_{-Fv(G)}K\wedge d \mid \langle
G,\emptyset,{\tt true}, \emptyset \rangle_1
\rightarrow^*_{\omega_t}
\langle \emptyset, \tilde K, d, T\rangle_n\not\rightarrow_{\omega_t}\}\\
\hspace{0.3cm}\cup\\
\hspace{0.3cm}\{{\tt false}\mid \langle G,\emptyset,{\tt true},
\emptyset \rangle_1 \rightarrow^*_{\omega_t} \langle G', \tilde K,
{\tt false}, T\rangle_n\}.
\end{array}
$$
\end{definition}

Analogously we can define the qualified answer of an annotated
program.

\begin{definition}\textsc{(Qualified answers for annotated programs)}.
Let $P$ be an annotated CHR program and let $G$ be a goal with $m$
CHR constraints. The set $\mathcal{QA'}_P(G)$ of qualified answers
for the query $G$ in the annotated program $P$ is defined as follows:
$$
\hspace*{-0.2cm}\begin{array}{l}
\mathcal{QA'}_P(G) = \\
\hspace*{0.3cm}\{\exists_{-Fv(G)}K\wedge d \mid \langle I_0^m (G),
{\tt true}, \emptyset\rangle_m\rightarrow^*_{\omega'_t}
\langle \tilde K, d, T\rangle_n\not\rightarrow_{\omega'_t}\}\\
\hspace{0.3cm}\cup\\
\hspace{0.3cm}\{{\tt false}\mid \langle I_0^m(G),
{\tt true},\emptyset\rangle_m\rightarrow^*_{\omega'_t}\langle
\tilde G', {\tt false}, T\rangle_n\}.
\end{array}
$$
\end{definition}

The following definition is introduced to describe the equivalence
of two intermediate states and it is used only in the proofs.
We consider two state equivalent when they are identical up to
renaming of local variables and renaming of identifiers and logical
equivalence of built-in constraints.

\begin{definition}[{\sc Inter-semantics State equivalence}]\label{def:PLQA}
 Let
 $\sigma=\la(H_1,C),\tilde H_2, D, T\ra_n \in {\it Conf_t}$ be a state in the transition system $\omega_t$ and let  $\sigma'=\la(\tilde K,C), D, T'\ra_m \in {\it Conf'_t}$ be a state in the transition system $\omega'_t$.

$\sigma$ and $\sigma'$ are \emph{equivalent} (and we write
$\sigma\equiv\sigma'$) if:
\begin{enumerate}
\item  there exist $\tilde K_1$ and $\tilde K_2$, such that $\tilde K =\tilde K_1\cup \tilde K_2$, $H_1 = chr(\tilde K_1)$ and $chr (\tilde H_2) = chr(\tilde K_2)$,
\item for each $l \in id (\tilde K_1)$, $l$ does not occur in $T'$,
\item there exists a renaming of identifier $\rho$ s.t. $T\rho=T'$ and $\tilde H_2 \rho= \tilde K_2$.
\end{enumerate}
\end{definition}

The following result shows the equivalence of the two introduced
semantics proving the equivalence (w.r.t. Definition \ref{def:PLQA})
of intermediate states. The proof
is easy by definition of $\omega_t$ and $\omega'_t$.

\begin{lemma}\label{lemma:intermequiv}
Let $P$ and $Ann(P)$ be respectively a CHR program and its annotated version.
Moreover, let
$\sigma \in {\it Conf_t}$ and let  $\sigma' \in {\it Conf'_t}$ such that
$\sigma\equiv\sigma'$.
Then the following holds
\begin{itemize}
  \item there exists a derivation $\delta = \sigma \rrarrow^*_{\omega_t} \sigma_1$ in $P$ if and only if
there exists a derivation $\delta' =\sigma' \rrarrow^*_{\omega'_t} \sigma'_1$ in $Ann(P)$ such  $\sigma_1\equiv\sigma'_1$
  \item the number of \textbf{Solve} (\textbf{Apply}) transition steps in $\delta$ and the number of \textbf{Solve'} (\textbf{Apply'}) transition steps in $\delta'$ are equal.
\end{itemize}
\end{lemma}
\textsc{Proof.}
 We show that any transition step from any state in one system can be imitated from a (possibly empty) sequence of  transition steps from an equivalent state in the other system to achieve an equivalent state.
 Moreover there exists a \textbf{Solve} (\textbf{Apply}) transition step in $\delta$ if and only if there exists a  \textbf{Solve'} (\textbf{Apply'}) transition step in $\delta'$.

 Then the proof follows by a straightforward inductive argument.

Let $\sigma=\la(H_1,C),\tilde H_2, D, T\ra_n \in {\it Conf_t}$
and let  $\sigma'=\la(\tilde K,C), D, T'\ra_m \in {\it Conf'_t}$ such that
$\sigma\equiv\sigma'$.

\begin{description}
\item[Solve and Solve':] they move a built-in constraint from the Goal store or the Store
respectively to the built-in constraint store. In this case let $C=C' \cup \{c\}$. By definition of the two transition systems
\[\begin{array}{l}
\sigma \rrarrow_{\omega_t}^{Solve}
\la(H_1,C'),\tilde H_2, D \wedge c, T\ra_n \mbox{ and }
\sigma ' \rrarrow_{\omega'_t}^{Solve'}
\la(\tilde K,C'), D \wedge c, T'\ra_m.
\end{array}
\]
By definition of $\equiv$, it is easy to check that $\la(H_1,C'),\tilde H_2, D \wedge c, T\ra_n \equiv
\la(\tilde K,C'), D \wedge c, T'\ra_m$.

\item[Introduce:] this kind of transition exists only in $\omega_t$ semantics and its
application labels a CHR constraint in the goal store and moves it in the CHR store.
In this case let $H_1=H'_1\uplus \{h\}$ and

$$
\begin{array}{l}
\sigma\rrarrow_{\omega_t}^{Introduce}
\la (H'_1, C),  \tilde H_2 \cup \{h\#n\}, D, T\ra_{n+1}.
\end{array}
$$
Let us denote $ \tilde H_2 \cup \{h\#n\}$ by $ \tilde H_2'$.
By definition of $\equiv$,  there exist $\tilde K_1$ and $\tilde K_2$, such that $\tilde K =\tilde K_1\cup \tilde K_2$, $H_1 = chr(\tilde K_1)$ and $chr (\tilde H_2) = chr(\tilde K_2)$. Therefore
there exists an identified atom $h\#m \in \tilde K_1$.
Let $n'=\rho(n)$ (where $n'=n$ if $n$ is not in the domain of $\rho$).
By construction and by hypothesis, $\tilde K'_1= \tilde K_1 \setminus \{h\#m\}$ and $\tilde K'_2= \tilde K_2\setminus \{h\#m\}$ are such that $\tilde K =\tilde K'_1\cup \tilde K'_2$, $H'_1 = chr(\tilde K'_1)$ and $chr (\tilde H'_2) = chr(\tilde K'_2)$.

Moreover, by definition of $\equiv$, for each $l \in id (\tilde K_1)$, $l$ does not occur in $T'$. Therefore, since by construction $\tilde K'_1 \subseteq \tilde K_1$, we have that for each $l \in id (\tilde K'_1)$, $l$ does not occur in $T'$.

Now, to prove that $\sigma' \equiv
\la (H'_1, C),  \tilde H_2', D, T\ra_{n+1}$, we have only to prove that there exists a renaming $\rho'$, such that
$T\rho'=T'$ and $\tilde H'_2 \rho'= \tilde K'_2$.

We can consider the new renaming $\rho'=\rho \circ \{n'/m, m/n'\}$. By definition $\rho'$ is a renaming of identifiers.
Since by construction, $m \not \in id (\tilde K_2)$, we have that if there exists $m'/m \in \rho$, then $m' \not \in id (\tilde H_2)$. Moreover, since $m \not \in id (\tilde K_2)$, if there is no $m/m' \in \rho$ then $m \not \in id (\tilde H_2)$.  By the previous observations, we have that $\tilde H'_2 \rho'= \tilde H_2 \rho
\cup \{h\#n\} \{n/m\} = \tilde K'_2$.
Finally, since $n$ does not occur in $T$, we have that $T \rho'=T\rho \{ m/n'\}=T'\{ m/n'\}$, where the last equality follows by hypothesis. Moreover  since $m \in id (\tilde K_1)$, we have that $m$ does  not occur in $T'$.
Therefore $T'\{ m/n'\}=T'$ and then the thesis.

\item[Apply and Apply':] Let $r@F'\backslash F'' \Leftrightarrow D_1 \,|\, B, C_1 \in P$ and
 $r@F'\backslash F'' \Leftrightarrow D_1 \,|\, \tilde B, C_1 \in Ann(P)$ be its annotated version which can
be applied to the considered state $\sigma'=\la(\tilde K,C), D, T'\ra_m$. In particular $F',F''$ match respectively with $\tilde P_1$ and
$\tilde P_2$. Without loss of generality, by using a suitable number of Introduce steps, we can assume that
$r@F'\backslash F'' \Leftrightarrow D_1 \,|\, B, C_1 \in P$ can be applied to $\sigma=\la(H_1,C),\tilde H_2, D, T\ra_n$. In particular, we can assume for $i=1,2$, there exists $\tilde Q_i \subseteq \tilde H_2$ such that $\tilde Q_i \rho = \tilde P_i$ and $F',F''$ match respectively with $\tilde Q_1$ and
$\tilde Q_2$.

By definition of $\equiv$, there exist $\tilde P_3$ and $\tilde Q_3$ such that
$\tilde Q_3 \rho = \tilde P_3$,
$\tilde K_2= \tilde P_1 \cup \tilde P_2\cup \tilde P_3$,
$\tilde H_2= \tilde Q_1 \cup \tilde Q_2 \cup \tilde Q_3$ and  let $x =
Fv(\tilde P_1, \tilde P_2)= Fv(\tilde Q_1, \tilde Q_2)$.

By construction, since $T \rho=T'$ and $(\tilde P_1, \tilde P_2)= (\tilde Q_1, \tilde Q_2)\rho$, we have that
\\
\begin{itemize}
\item $r@id(\tilde P_1,\tilde P_2) \not \in T'$ if and only if
$r@id(\tilde Q_1,\tilde Q_2) \not \in T$ and
\item $CT \models D \rightarrow
\exists _{x}(( (F', F'')=chr (\tilde P_1, \tilde P_2))\wedge D_1)$ if and only if
$CT \models D \rightarrow
\exists _{x}( ((F', F'')=chr (\tilde Q_1, \tilde Q_2))\wedge D_1)$.
\end{itemize}

Therefore, by definition of {\bf Apply} and of {\bf Apply'}
\[\sigma \rrarrow^{Apply}_{\omega_t} \la \{H_1,C\}\uplus \{B, C_1 \},(\tilde Q_1, \tilde Q_3),
((F',  F'')=chr(\tilde Q_1, \tilde Q_2))\wedge D, T_1 \ra_n\]
if and only if
\[\sigma' \rightarrow^{Apply'}_{\omega'_t}  \la(\tilde K_1,\tilde P_1, \tilde P_3 ,C, \tilde B', C_1),
((F',  F'')=chr(\tilde P_1, \tilde P_2))\wedge D, T'_1 \ra_o\]
where
\begin{itemize}
  \item $T'=T \cup \{{\it r}@id (\tilde Q_1)\}$ if $\tilde Q_2=\emptyset$,
 otherwise $T_1=T$,
  \item $(\tilde B',\emptyset ,o )= inst(\tilde B,\emptyset ,m)$ and
  \item $T_1'=T'\cup \{r @id (\tilde P_1)\}$ if $\tilde Q_2=\emptyset$,
 otherwise $T_1'=T'$.
\end{itemize}

Let $\sigma_1= \la \{H_1,C\}\uplus \{B, C_1 \},(\tilde Q_1, \tilde Q_3),
((F',  F'')=chr(\tilde Q_1, \tilde Q_2))\wedge D, T_1 \ra_n$ and
$\sigma'_1= \la(\tilde K_1,\tilde P_1, \tilde P_3 , \tilde B',C, C_1),
((F',  F'')=chr(\tilde P_1, \tilde P_2))\wedge D, T'_1 \ra_o$. \\

Now, to prove the thesis, we have to prove that
$\sigma_1 \equiv \sigma'_1$.

The following holds.

\begin{enumerate}
\item  There exist $\tilde K'_1= (\tilde K_1, \tilde B') $ and $\tilde K'_2=(\tilde P_1, \tilde P_3)$, such that $(\tilde K_1,\tilde P_1, \tilde P_3 , \tilde B') =\tilde K'_1\cup \tilde K'_2$,
    $H_1 \uplus B = chr(\tilde K'_1)$ and $chr (\tilde Q_1,\tilde Q_3) = chr(\tilde K'_2)$.
\item Since for each $l \in id (\tilde K_1)$, $l$ does not occur in $T'$,
 $\tilde P_1 \subseteq \tilde K_2$ and by definition of \textbf{Apply'} transition, we have that for each $l \in id (\tilde K'_1)= id(\tilde K_1, \tilde B')$, $l$ does not occur in $T'_1$,
\item By construction and since $T\rho=T'$, we have that $T_1\rho=T'_1$.
Moreover, by construction  $(\tilde Q_1,\tilde Q_3) \rho= (\tilde P_1,\tilde P_3)= \tilde K'_2$.
\end{enumerate}
By definition, we have that $\sigma_1 \equiv \sigma'_1$ and then the thesis.
\end{description}
\noindent{$\Box$}

\begin{proposition}\label{lemma:nequality}
Let $P$ and $Ann(P)$ be respectively a CHR program and its annotated version.
Then, for every goal $G$,
$$\mathcal{QA}_{P}(G) = \mathcal{QA'}_{Ann(P)}(G)$$
holds.
\end{proposition}
\textsc{Proof.} By definition of $\mathcal{QA}$ and of $\mathcal{QA'}$,
the initial states of the two transition system are equivalent. Then the proof follows by Lemma \ref{lemma:intermequiv}.

\section{The unfolding rule}\label{sec:unfolding}

In this section we define the \emph{unfold operation} for CHR
simpagation rules. As a particular case we obtain also
unfolding for  simplification and propagation rules, as these can
be seen as particular cases of the former.

The unfolding allows to replace  a  conjunction $S$ of constraints
(which can be seen as a procedure call) in the body of a rule $r$
by the body of a rule  $v$,  provided that the head of $v$ matches
with $S$, by assuming the built-in constraints in
the guard and in the body of the rule $r$. More precisely, assume that
the built-in constraints in the guard and in the body
of the rule $r$ imply that the head $H$ of $v$,
instantiated by a substitution $\theta$, matches with the
conjunction $S$  (in the body of  $r$). Then the unfolded rule is
obtained from $r$ by performing the following steps:  1) the new
guard in the unfolded rule is the conjunction of the guard of $r$
with the guard of $v$, the latter instantiated by $\theta$ and
without  those constraints that are entailed by the built-in
constraints which are in $r$; 2) the body of $v$ and the equality
$H = S$ are added to the body of $r$; 3) the conjunction of constraints $S$ can
be removed, partially removed or left in the body of  the unfolded
rule, depending on the fact that $v$ is a simplification, a
simpagation or a propagation rule, respectively; 4) as for the
local token store  $T_r$ associated to every rule $r$, this is
updated consistently during the unfolding operations in order to
avoid that a propagation rule is used twice to unfold the same
sequence of constraints.

Before formally defining the unfolding we need to define the
function \[clean: Goals \times Token \rightarrow Token,\] as
follows: $clean (\tilde B,T)$ deletes from $T$ all the tokens for
which at least one identifier is not present in the identified
goal $\tilde B$. More formally
$$\begin{array}{l}
        clean(\tilde B, T)= \{t\in T\mid  t=r@i_1, \ldots, i_k \mbox{ and }
        i_j\in id(\tilde B), \mbox{ for each } j \in [1,k]\}.
\end{array}
$$
Recall also that we defined  {\em chr(h$\#$i)=h}.

\begin{definition}\textsc{(Unfold).}\label{def:unf}
Let $P$ be an annotated CHR program  and let $r, sp\in P$ be two
annotated rules such that:
$$
\begin{array}{rcl}
r@H_1\backslash H_2 &\Leftrightarrow&  D\,|\,\tilde K, \tilde S_1, \tilde S_2, C; T \mbox{ and}\\
v@H_1'\backslash H_2' &\Leftrightarrow & D '\,|\, \tilde B;
T'
\end{array}
$$
where $C$ is the conjunction of all the built-in constraints in the body
of $r$ and $CT \models (C \wedge D) \rightarrow chr(\tilde S_1, \tilde S_2)= (H_1',
H_2')\theta$, that is, the constraints $H'_1$  in the head of rule
$v\, $ match with $chr(\tilde S_1)$ and $H_2'$ matches with $chr(\tilde
S_2)$ by using the substitution $\theta$,
once the built-in constraints in $r$ are assumed. Furthermore assume that
$m$ is the greatest identifier which appears in the
rule $r$ and that $(\tilde B_1, T_1, m_1)=inst(\tilde B, T',m)$.
Then the \emph{unfolded} rule is:
$$r@ H_1\backslash H_2
\Leftrightarrow D, (D''\theta)\, |\, \tilde K,\tilde S_1,\tilde
B_1,C, chr(\tilde S_1, \tilde S_2)= (H_1', H_2'); T''$$ where $v @id (\tilde
S_1, \tilde S_2) \not \in T$,  $V\subseteq
D'$ is the greatest set of built-in constraints $c$, such that $CT \models C\wedge D\rightarrow c\theta$,
$D''= D'\backslash V$, the
constraint $(D, (D''\theta))$ is satisfiable and
\begin{itemize}
    \item if $H_2'=\emptyset$ then $T''=clean((\tilde K,\tilde S_1) , T) \cup T_1 \cup\{v @id (\tilde
S_1)\}$
    \item if $H_2'\not =\emptyset$ then $T''=clean((\tilde K,\tilde S_1) , T) \cup T_1$.
    \end{itemize}
\end{definition}

Note that we use  the function $inst$  (Definition~\ref{definst})
in order to increment the value of the
identifiers associated to atoms in the unfolded rule. This allows
us to distinguish the new identifiers introduced in the unfolded
rule from the old ones. Note also that the condition on the token
store is needed to obtain a correct rule.  Consider for example a ground annotated
program  $P=\{r_1@ h \Leftrightarrow \tilde k, \, r_2@k
\Rightarrow \tilde s, \, r_3@s,s\Leftrightarrow \tilde B\}$ and let
$h$ be  the start goal. In this case the unfolding could change
the semantics if the token store were not used. In fact, according
to the semantics proposed in Table \ref{omega-t} or
\ref{tab:operational-semantics}, we have the following
computation: $\tilde h\rightarrow^{(r_1)}\tilde
k\rightarrow^{(r_2)}\tilde k, \tilde
s\not\rightarrow_{\omega_t}$. On the other hand,  considering an
unfolding without the update of  the token store one would have
$r_1@ h\Leftrightarrow \tilde k \stackrel{\mbox{\tiny{unfold using
$r_2$}}}{\longrightarrow} r_1@ h\Leftrightarrow \tilde k, \tilde s
\stackrel{\mbox{\tiny{unfold using
$r_2$}}}{\longrightarrow}\sout{r_1@h\Leftrightarrow \tilde k,
\tilde s, \tilde s}\stackrel{\mbox{\tiny{unfold using
$r_3$}}}{\longrightarrow}r_1@h\Leftrightarrow \tilde k, \tilde B$
so, starting from the constraint $h$ we could arrive to constraint $k, B$, that
is not possible in the original program (the clause obtained after
the wrongly applied unfolding rule is underlined).

As previously mentioned, the unfolding rules for simplification
and propagation can be obtained as particular cases of
Definition~\ref{def:unf}, by setting $H_1'=\emptyset$ and  $H_2'
=\emptyset$, respectively, and by considering accordingly the
resulting unfolded rule. In the following examples we will use
$\odot$ to denote both  $\Leftrightarrow$ and $\Rightarrow$.

\begin{example}\label{ex:gen_adam}
The following program $P=\{r_1, r_2, \bar r_2\}$  deduces
information  about genealogy. Predicate $f$ is considered as
father, $g$ as grandfather, $gs$ as grandson and $gg$ as
great-grandfather. The following rules are such that we can unfold
some constraints in the body of $r_1$ using the rule $r_2$ $[\bar
r_2]$.
$$
\begin{array}{l}
r_1@f(X, Y), f(Y, Z), f(Z, W)\odot g(X, Z)\#1,f(Z, W)\#2,gs(Z, X)\#3.\\
r_2@g(X, Y), f(Y, Z)\odot gg(X, Z)\#1.\\
\bar r_2@g(X, Y)\backslash f(Y, Z) \Leftrightarrow gg(X, Z)\#1.
\end{array}
$$
Now we  unfold the body of rule $r_1$ by using the rule
$r_2$ where we assume $\odot=\Leftrightarrow$ (so we have a simplification rule).  We use
$inst(gg(X,Z)\#1,\emptyset,3) = (gg(X,Z)\#4,\emptyset,4)$ and a renamed version of $r_2$
$$r_2@g(X', Y'), f(Y', Z') \Leftrightarrow gg(X', Z')\#1.$$
 in order to avoid variable clashes.
So the new unfolded rule is:
$$
\begin{array}{l}
r_1@f(X, Y),f(Y, Z), f(Z, W)\odot  gg(X', Z')\#4,
gs(Z,X)\#3, X'=X, Y'=Z, Z'=W.
\end{array}
$$

Now, we unfold the body of rule $r_1$ by using the simpagation
rule $\bar r_2$. As before,
\[inst(gg(X,Z)\#1,\emptyset,3) = (gg(X,Z)\#4,\emptyset,4)\] and a renamed version of $\bar
r_2$
$$\bar r_2@g(X', Y')\backslash f(Y', Z') \Leftrightarrow gg(X', Z')\#1.$$
is used to avoid variable clashes. The new unfolded rule is:
$$
\begin{array}{l}
r_1@f(X, Y),f(Y, Z), f(Z, W)\odot  g(X, Z)\#1,\\
\hspace{1cm} gg(X', Z')\#4,gs(Z,X)\#3, X'=X, Y'=Z, Z'=W.
\end{array}
$$

Finally we unfold the body of $r_1$ by using the $r_2$ rule where $\odot$ = $\Rightarrow$ is assumed (so we have a propagation rule).
As usual, $inst(gg(X,Z)\#1,\emptyset,3) = (gg(X,Z)\#4,\emptyset,4)$ and a renamed version of $r_2$ is used to avoid variable clashes:
$$r_2@g(X', Y'), f(Y', Z') \Rightarrow gg(X', Z')\#1.$$
and so the new unfolded rule is:
$$
\begin{array}{l}
r_1@f(X, Y),f(Y, Z), f(Z, W)\odot g(X, Z)\#1,\\
\hspace{1cm}f(Z, W)\#2, gs(Z, X)\#3, gg(X', Z')\#4, X'=X,
Y'=Z,Z'=W; \{r_2@1, 2\}.
\end{array}
$$
\end{example}

The following example considers more specialized rules  with guards which are not $\texttt{true}$.

\begin{example}\label{ex:gen_adam_refined}
The following  program $P=\{r_1, r_2, \bar r_2\}$ specializes the rules introduced in Example \ref{ex:gen_adam}
to the genealogy of Adam. So here we remember that Adam was father of Seth; Seth was father
of Enosh; Enosh was father of Kenan. As before, we consider the predicate $f$ as father,
$g$ as grandfather, $gs$ as grandson and $gg$ as great-grandfather.
$$
\begin{array}{l}
r_1@f(X, Y), f(Y, Z), f(Z,W)\odot X= Adam, Y=Seth\,|\,\\
\hspace{1cm}g(X, Z)\#1,f(Z, W)\#2, gs(Z, X)\#3, Z=Enosh.\\
r_2@g(X, Y), f(Y, Z)  \odot  X=Adam, Y=Enosh\,|\, gg(X, Z)\#1, Z=Kenan.\\
\bar r_2@g(X, Y) \backslash f(Y, Z) \Leftrightarrow  X=Adam, Y=Enosh\,|\, gg(X, Z)\#1,Z=Kenan.
\end{array}
$$

If we unfold $r_1$ by using (a suitable renamed version of) $r_2$,
where we assume $\odot=\Leftrightarrow$, we obtain:
$$
\begin{array}{l}
r_1@f(X, Y),f(Y, Z)f(Z, W)\odot X=Adam, Y=Seth\,|\,gg(X', Z')\#4, Z'= Kenan, \\
\hspace{1cm}gs(Z, X)\#3, Z=Enosh, X'=X, Y'=Z, Z'=W.
\end{array}
$$
When $\bar r_2$ is considered to unfold $r_1$ we have
$$
\begin{array}{l}
r_1@f(X, Y),f(Y, Z)f(Z, W)\odot X=Adam, Y=Seth\,|\,g(X, Z)\#1,gg(X', Z')\#4,   \\
\hspace{1cm}Z'= Kenan, gs(Z, X)\#3, Z=Enosh, X'=X, Y'=Z, Z'=W.
\end{array}
$$
Finally if we assume $\odot=\Rightarrow$ in $r_2$ from the unfolding we obtain
$$
\begin{array}{l}
r_1@f(X, Y),f(Y, Z), f(Z, W)\odot X=Adam, Y=Seth\,|\,g(X, Z)\#1, f(Z, W)\#2,\\
\hspace{1cm}gs(Z, X)\#3, gg(X', Z')\#4, Z'= Kenan, Z=Enosh, X'=X, Y'=Z,\\
\hspace{1cm} Z'=W; \{r_2@1, 2\}.
\end{array}
$$

Note that  $X'=Adam, Y'=Enosh$ are not
added to the guard of the unfolded rule  because $X'=Adam$ is
entailed by the guard of $r_1$ and $Y'=Enosh$ is entailed
by the built-in constraints in the body of $r_1$.
\end{example}

We prove now the correctness of our unfolding definition.
Before the introduction of the proposition which proves the correctness of our unfolding, three new
definitions are given. The first one presents the concept of built-in
free state. Said state either has no built-in constraints in the first component
or the built-in store is unsatisfiable.

\begin{definition}[{\sc Built-in free State}]\label{def:BFS}
Let  $\sigma=\la G, \tilde S , D, T\ra_o\in {\it Conf_t}$ ($\sigma=\la\tilde G , D, T\ra_o\in {\it Conf'_t}$).
The state $\sigma$ is built-in free  if either $D=\tt false$  or
$G$ ($\tilde G$) is a multiset of (identified) CHR-constraints.
\end{definition}

The second definition introduces the state equivalence between states in ${\it Conf'_{t}}$.
Note that in such definition, the
equivalence operator is represented with the symbol $\simeq$.

\begin{definition}[{\sc State equivalence}]\label{def:SE}
Let  $\sigma=\la\tilde G , D, T\ra_o$ and  $\sigma'=\la \tilde G', D', T'\ra_o$ be states in ${\it Conf'_t}$.
$\sigma$ and $\sigma'$ are equivalent and we write $\sigma\simeq\sigma'$ if one of the following facts hold.
\begin{itemize}
  \item either $D=\tt false$ and $D'=\tt false$
  \item or $\tilde G =\tilde G'$, $CT\models D \leftrightarrow D'$ and $clean(\tilde G, T)= clean(\tilde G', T)$.
\end{itemize}
\end{definition}

Finally the third definition presents the normal derivation. A derivation
is called normal if no other \textbf{Solve} (\textbf{Solve'}) transition are possible
when an \textbf{Apply} (\textbf{Apply'}) one happens.

\begin{definition}[{\sc Normal derivation}]\label{def:ND}
Let  $P$ be a (possibly annotated) CHR program  and let
$\delta$ be a derivation in $P$.
We say that $\delta$ is normal if it uses a transition
\textbf{Solve} (\textbf{Solve'}) as soon as possible, namely it is possible to use a transition
\textbf{Apply} (\textbf{Apply'})
on a state $\sigma$ only if $\sigma$ is built-in free.
\end{definition}

Note that, by definition, given a CHR program $P$, $\mathcal{QA}(P)$ can be calculated by
considering only normal derivations. Analogously for an annotated CHR program $P'$.
The proof  of the following proposition is straightforward and hence it is omitted.

\begin{proposition}\label{prop:solonorm}
Let $P$ be CHR program  and let $P'$ an annotated CHR program.
Then
$$
\begin{array}{lcl}
\mathcal{QA}_P(G) &=&
\{\exists_{-Fv(G)}K\wedge d \mid
\delta= \langle
G,\emptyset,{\tt true}, \emptyset \rangle_1
\rightarrow^*_{\omega_t}
\langle \emptyset, \tilde K, d, T\rangle_n\not\rightarrow_{\omega_t}\\
&&\hspace{0.6cm}\mbox{and $\delta$ is normal}\}\\
&&\cup\\
&&\{{\tt false}\mid \delta = \langle G,\emptyset,{\tt true},
\emptyset \rangle_1 \rightarrow^*_{\omega_t} \langle G', \tilde K,
{\tt false}, T\rangle_n\\
&&\hspace{0.6cm}\mbox{and $\delta$ is normal}\}
\end{array}
$$
and
$$\begin{array}{lcl}
\mathcal{QA'}_P(G) &=&
\{\exists_{-Fv(G)}K\wedge d \mid
\delta=\langle I_0^m (G),
{\tt true}, \emptyset\rangle_m\rightarrow^{*}_{\omega'_t}
\langle \tilde K, d, T\rangle_n\not\rightarrow_{\omega'_t}\\
&&\hspace{0.6cm}\mbox{and $\delta$ is normal}\}\\
&& \cup\\
&&\{{\tt false}\mid
\delta=\langle I_0^m(G),
{\tt true},\emptyset\rangle_m\rightarrow^{*}_{\omega'_t}\langle
\tilde G', {\tt false}, T\rangle_n\\
&&\hspace{0.6cm}\mbox{and $\delta$ is normal}\}.
\end{array}
$$
\end{proposition}

\begin{proposition}\label{prop:servequality}
Let $r, v$ be annotated CHR rules and $r'$  be the result
of the unfolding of $r$ with respect to $v$. Let $\sigma$ be a generic built-in free state such that we can use the transition \textbf{Apply'} with the clause $r'$ obtaining the state $\sigma_{r'}$ and then the built-in free state $\sigma_{r'}^f$. Then we can construct a derivation which uses at most the clauses $r$ and $v$ and obtain a built-in free state $\sigma^f$ such that $\sigma_{r'}^f \simeq \sigma^f$.
\end{proposition}

\textsc{Proof.}
Assume that
$$
\begin{array}{rl}
\sigma&\rrarrow^{r'}\sigma_{r'}\rrarrow^{Solve^{*}}
\sigma_{r'}^f\\
&\searrow_{\, r}
\sigma_r \rrarrow^{Solve^{*}}
\sigma_r^f (\rrarrow^v
\sigma_v\rrarrow^{Solve^{*}}
\sigma_v^f)
\end{array}
$$

The labeled arrow $\rrarrow^{Solve^{*}}$ means that only solve transitions are applied.
Moreover
\begin{itemize}
           \item if $\sigma_r^f$ has the form
$\la \tilde G, {\tt false}, T\ra$ then the derivation between the parenthesis is not present and $\sigma^f=\sigma_r^f$.
           \item the derivation between the parenthesis is present and $\sigma^f=\sigma_v^f$, otherwise.
         \end{itemize}

\noindent{\textbf{Preliminaries}:}
Let  $\sigma=\la (\tilde H_1,\tilde H_2,\tilde H_3) , C, T\ra_j$ be a built-in free state
and let
$r@H'_1\backslash H'_2 \Leftrightarrow  D_r\,|\,\tilde K, \tilde S_1, \tilde S_2, C_r; T_r $ and
$v@S_1'\backslash S_2' \Leftrightarrow  D_v\,|\, \tilde P, C_v;T_v$
where $C_r$ is the conjunction of all the built-in constraints in the body
of $r$ and
\begin{equation}\label{13marzo1}
    CT \models (D_r \wedge C_r) \rightarrow chr(\tilde S_1, \tilde S_2)=(S_1',
S_2')\theta.
\end{equation}
Furthermore assume that $m$ is the greatest identifier which appears in the
rule $r$ and that $inst(\tilde P, T_v,m)=(\tilde P_1, T_1, m_1)$.
Then the \emph{unfolded} rule is:
\[
    r'@ H'_1\backslash H'_2
\Leftrightarrow D_r, (D_v'\theta)\, |\, \tilde K,\tilde S_1,\tilde
P_1, C_r, C_v, chr(\tilde S_1, \tilde S_2)= (S_1', S_2'); T_{r'}
\]
where $v @id (\tilde S_1, \tilde S_2) \not \in T_r$,
$V\subseteq D_v$ is the greatest set of built-in constraints $c$,
such that
$CT \models (D_r\wedge C_r)\rightarrow c\theta,$
$D_v'= D_v\backslash V$, the
constraint $(D_r, (D_v'\theta))$ is satisfiable and
\begin{itemize}
    \item if $S_2'=\emptyset$ then $T_{r'}=clean((\tilde K,\tilde S_1) ,  T_r) \cup T_1 \cup
              \{v @id (\tilde S_1)\}$
    \item if $S_2'\not =\emptyset$ then $T_{r'}=clean((\tilde K,\tilde S_1) ,  T_r) \cup T_1$.
    \end{itemize}
    By previous observations, we have that
    \begin{equation}\label{10dic2}
    CT \models (D_r\wedge C_r)\rightarrow V\theta.
\end{equation}
\noindent{\textbf{The proof}:} By definition of the transition \textbf{Apply'}, we have that
\begin{equation}\label{10dic1}
    CT\models C\rightarrow \exists_x((chr(\tilde H_1,
\tilde H_2)=(H'_1, H'_2))\wedge D_r\wedge (D_v'\theta)),
\end{equation}
where $ x=Fv(H'_1,H'_2)$ and
$$\begin{array}{l}
    \sigma_{r'}=\la (\tilde Q, C_r, C_v,chr(\tilde S_1, \tilde S_2)= (S_1', S_2')),
chr(\tilde H_1, \tilde H_2)= (H_1', H_2')\wedge C, T_3\ra_{j+m_1},
  \end{array}
  $$
     where
     $\tilde Q=(\tilde H_1,\tilde H_3,\tilde Q_1)$, with
     $inst ((\tilde K,\tilde S_1,\tilde P_1), T_{r'}, j)=(\tilde Q_1, T_{r'}', j+m_1)$ and
\begin{itemize}
    \item if $H_2'=\emptyset$ then $T_3=T \cup T_{r'}' \cup\{r @id (\tilde
H_1)\}$
    \item if $H_2'\not =\emptyset$ then $T_3=T \cup T_{r'}'$.
    \end{itemize}
    Therefore, by definition
    $$
    \sigma_{r'}^f=\la \tilde Q, C_{r'}^f, \, T_3\ra_{j+m_1}.$$
    where
    $$\begin{array}{l}
       CT \models C_{r'}^f \leftrightarrow  C_r\wedge  C_v\wedge chr(\tilde S_1, \tilde S_2)= (S_1', S_2') \wedge chr(\tilde H_1, \tilde H_2)= (H_1', H_2')\wedge C.
     \end{array}
    $$

On the other hand, since by (\ref{10dic1}),
\[CT\models C\rightarrow \exists_x((chr(\tilde H_1,
\tilde H_2)=(H'_1, H'_2))\wedge D_r)
\]
by definition of the transition \textbf{Apply'}, we have that
$$\begin{array}{l}
    \sigma_{r}=\la (\tilde Q_2, C_r),
          chr(\tilde H_1, \tilde H_2)= (H_1', H_2')\wedge C, T_4\ra_{j+m},
  \end{array}
  $$

     where $\tilde Q_2=(\tilde H_1,\tilde H_3,\tilde K'',\tilde S''_1,\tilde S''_2)$,\\
     $((\tilde K'',\tilde S''_1,\tilde S''_2), T_2, j+m) = inst ((\tilde K,\tilde S_1,\tilde S_2), T_r, j)$ and
\begin{itemize}
    \item if $H_2'=\emptyset$ then $T_4=T \cup T_2 \cup\{r @id (\tilde
H_1)\}$
    \item if $H_2'\not =\emptyset$ then $T_4=T \cup T_2$.
    \end{itemize}
    Therefore, by definition
    $$ \sigma_{r}^f=\la\tilde Q_2, C_{r}^f, \, T_4\ra_{j+m}.
  $$
  where
\begin{equation}\label{13marzo2}
CT \models C_{r}^f \leftrightarrow  C_r\wedge
     chr(\tilde H_1, \tilde H_2)= (H_1', H_2')\wedge C.
\end{equation}

Now, we have two possibilities
\begin{description}
  \item[($C_{r}^f = \tt false$).] In this case, by construction we have that $C_{r'}^f = \tt false$. Therefore $ \sigma_{r'}^f \simeq  \sigma_{r}^f$ and then the thesis.
  \item[($C_{r}^f    \neq \tt false$).]
  By definition, since
$chr(\tilde S_1, \tilde S_2)=chr(\tilde S''_1, \tilde S''_2)$, by (\ref{13marzo1}), (\ref{10dic2}) and (\ref{10dic1}), we have that
\[\begin{array}{ll}
    CT\models  & (C_r\wedge
     chr(\tilde H_1, \tilde H_2)= (H_1', H_2')\wedge C)
     \rightarrow  \\
     & (\exists_y((chr(\tilde S_1,
\tilde S_2)=(S'_1, S'_2))\wedge D_v)),
  \end{array}
 \]
where $ y=Fv(S'_1, S'_2)$. Therefore
by (\ref{13marzo2})
\[CT\models  C_{r}^f
     \rightarrow  (\exists_y((chr(\tilde S_1,
\tilde S_2)=(S'_1, S'_2))\wedge D_v))
 \]

and
$$\begin{array}{ll}
    \sigma_{v}=&\la (Q_3 ,  C_v),
     chr(\tilde S_1, \tilde S_2)= (S_1', S_2')\wedge C_r\wedge\\
     &\hspace{0.5cm}chr(\tilde H_1, \tilde H_2)= (H_1', H_2')\wedge C, \, T_5\ra_{m_1},
  \end{array}
  $$
     where
     $\tilde Q_3=(\tilde H_1,\tilde H_3,\tilde K'',\tilde S''_1, \tilde P_2)$, with
     $inst (\tilde P, T_v, j+m) = (\tilde P_2, T_v', m_1)$ and
\begin{itemize}
    \item if $S_2'=\emptyset$ then $T_5=T_4 \cup T_v' \cup\{v @id (\tilde
S''_1)\}$
    \item if $S_2'\not =\emptyset$ then $T_5=T_4 \cup T_v'$.
    \end{itemize}
    Finally by definition, we have that
$$
    \sigma_{v}^f=\la \tilde Q_3,C_{v}^f, \, T_5\ra_{m_1},
$$
  where
  $$\begin{array}{l}
    C_{v}^f \leftrightarrow C_v \wedge
     chr(\tilde S_1, \tilde S_2)= (S_1', S_2')\wedge C_r\wedge
     chr(\tilde H_1, \tilde H_2)= (H_1', H_2')\wedge C.
  \end{array}
  $$

  If $C_{v}^f =\tt false $ then the proof is analogous to the previous case and hence it is omitted.
  Otherwise, observe that by construction,
$\tilde Q=(\tilde H_1,\tilde H_3,\tilde Q_1)$, where
$\tilde Q_1$ is obtained from $(\tilde K,\tilde S_1,\tilde P_1)$ by adding the natural $j$ to each identifier in
$(\tilde K,\tilde S_1)$ and by adding the natural $j+m$ to each identifier in
$\tilde P$.
Analogously, by construction,
$\tilde Q_3=(\tilde H_1,\tilde H_3,\tilde K'',\tilde S''_1, \tilde P_2)$, where
$(\tilde K'',\tilde S''_1)$  are obtained from $(\tilde K,\tilde S_1)$ by adding the natural $j$ to each identifier in
$(\tilde K,\tilde S_1)$ and $\tilde P_2$ is obtained from $\tilde P$ by adding the natural  $j+m$ to each identifier in $\tilde P$.

Therefore $\tilde Q=\tilde Q_3$ and then, to prove the thesis, we have only to prove that \[clean(\tilde Q,T_3)=clean(\tilde Q, T_5).\]

Let us introduce the function $inst': \{Token\}\times\mathbb{N} \rrarrow \mathbb{N}$ as
the restriction of the function $inst$ to token sets and natural numbers, namely
$inst'(T,n)= T'$, where $T'$ is obtained from $T$ by
incrementing each identifier in $T$ with $n$. So, since
$T'_2= inst'(T_2, j)$,
$T_2= clean((\tilde K, \tilde S_1), T_r)\cup T_1\cup \{v@id(\tilde S_1)\mid \mbox{if } S_2= \emptyset\}$ and
$T_1=inst'(T_v, m)$, we have that

$$
\begin{array}{lcl}
T_3 & = & T\, \cup \, T'_2\, \cup \,\{r@id(\tilde H_1)\mid \mbox{if } H_2= \emptyset\}\\
       &=& T \, \cup \, inst'(clean((\tilde K, \tilde S_1), T_r), j)\, \cup \, inst'(T_v, j+ m)
       \,\cup \\
       && inst'(\{v@id(\tilde S_1) \mid \mbox{if } S_2= \emptyset\}, j)\, \cup \,
       \{r@id(\tilde H_1)\mid \mbox{if } H_2= \emptyset\}
\end{array}
$$
Analogously, $T_4= T \cup T'_r \cup \{r@id(\tilde H_1)\mid \mbox{if } H_2= \emptyset\}$,
$T'_r = inst'(T_r, j)$ and $T'_v= inst'(T_v,j+m)$, we have that
$$
\begin{array}{lcl}
T_5 &=& T_4\, \cup \, T'_v \, \cup \, \{v@id(\tilde S''_1)\mid \mbox{if } S''_2= \emptyset\}\\

    &=& T \, \cup \, inst'(T_r, j) \, \cup \, \{r@id(\tilde H_1)\mid \mbox{if } H_2= \emptyset\}\,
    \cup \, inst'(T_v,j+m) \, \cup \, \\
    &&\{v@id(S''_1)\mid \mbox{if } S''_2= \emptyset\}
    \end{array}
$$
Now, since by construction $(S''_1, S''_2)$ is obtained from $(S_1,  S_2)$ by adding the natural $j$ to each identifier, we have that $inst'(\{v@id(\tilde S_1) \mid \mbox{if } S_2= \emptyset\}, j)= \{v@id(\tilde S''_1)\mid \mbox{if } S''_2= \emptyset\}$.
Moreover, by definition of annotated rule $id(T_r) \subseteq id (\tilde K, \tilde S_1, \tilde S_2)$
and $\tilde Q=(\tilde H_1,\tilde H_3,\tilde Q_1)$, where
$\tilde Q_1$ is obtained from $(\tilde K,\tilde S_1,\tilde P_1)$ by adding the natural $j$ to each identifier in
$(\tilde K,\tilde S_1)$ and by adding the natural $j+m$ to each identifier in
$\tilde P$. Then $clean(\tilde Q, inst'(clean((\tilde K, \tilde S_1), T_r), j))=
clean(\tilde Q, inst'(T_r, j))$ and then the thesis.
\end{description}
\noindent{$\Box$}

We prove now the correctness of our unfolding rule.

\begin{proposition}\label{lemma:equality}
Let $P$ be an annotated CHR program with $r, v\in P$. Let $r'$  be the result
of the unfolding of $r$ with respect to $v$ and let $P'$ be the program
obtained from $P$ by adding rule $r'$. Then, for every goal $G$,
$\mathcal{QA'}_{P'}(G) = \mathcal{QA'}_P(G)$ holds.
\end{proposition}

\textsc{Proof.}
We prove the two inclusions separately.
\begin{description}
  \item[{\bf ($\mathcal{QA'}_{P'}(G) \subseteq \mathcal{QA'}_P(G)$)}] The proof follows from Propositions \ref{prop:solonorm} and \ref{prop:servequality} and by a straightforward inductive argument.
  \item[{\bf ($\mathcal{QA'}_{P}(G) \subseteq \mathcal{QA'}_{P'}(G)$)}]
  The proof is by contradiction. Assume that there exists $(K'\wedge d') \in \mathcal{QA'}_{P}(G) \setminus \mathcal{QA'}_{P'}(G)$. By definition there exists a derivation
  \[\delta=\langle I_0^m (G),
{\tt true}, \emptyset\rangle_m\rightarrow^{*}_{\omega'_t}
\langle \tilde K, d, T\rangle_n\not\rightarrow_{\omega'_t}\] in $P$, such that $(K'\wedge d') =
\exists _{-Fv(G)}(chr( \tilde K)\wedge d)$. Since $P \subseteq P'$, we have that there exists the derivation
$\langle I_0^m (G),
{\tt true}, \emptyset\rangle_m\rightarrow^{*}_{\omega'_t}
\langle \tilde K, d, T\rangle_n$ in $P'$. Moreover, since $P' =P \cup \{r'\}$ and by hypothesis $(K'\wedge d')\not  \in \mathcal{QA'}_{P'}(G)$, we have that there exists a derivation step $\langle \tilde K, d, T\rangle_n\rightarrow_{\omega'_t}
\langle \tilde K_1, d_1, T_1\rangle_{n_1}$ by using the clause $r'$.
Then, by definition of unfolding there exists a derivation step $\langle \tilde K, d, T\rangle_n\rightarrow_{\omega'_t}
\langle \tilde K_2, d_2, T_2\rangle_{n_2}$ in $P$, by using the clause $r$ and then we have a contradiction.
\end{description}

\noindent{$\Box$}\\

\section{Safe rule replacement}\label{sec:safty-rule-deletion}
Previous corollary shows that we can safely add to a program $P$
a rule resulting from the unfolding, while preserving the
semantics of  $P$ (in terms of qualified answers).
However, when a rule $r$ in program $P$ has been unfolded producing
the new rule $r'$, in some cases we would like also to replace $r$ by $r'$ in $P$,
since this could improve the efficiency of the resulting program.
Performing such a replacement while preserving the semantics
is in general a very difficult task for three reasons.

First of all, anticipating the guard of $v$ in the guard of $r$
(as we do in the unfold operation) could lead to loose some
computations  when the unfolded rule $r'$  is used rather than the
original rule $r$. This is shown by the following example.

\begin{example}\label{esempio:mau}
Let us consider the program
\[\begin{array}{rlll}
  P= \{ &r  @ p(Y)  \Leftrightarrow q(Y).\\
  & r' @ q(Z) \Leftrightarrow  Z=a \,|\, .\}\\
\end{array}
\]
where we do not consider the identifiers (and the local token store) in
the body of rules, because we do not have propagation rules in
$P$.

The unfolding  of $r$ by using the rule $r'$ returns the new
rule $r @ p(Y) \Leftrightarrow Y=a  \,|\,  Y=Z$. The
program
\[\begin{array}{rlll}
  P'= \{ &r @ p(Y)  \Leftrightarrow Y=a  \,|\,  Y=Z.\\
 & r' @ q(Z) \Leftrightarrow  Z=a  \,|\,. \}\\
\end{array}
\]
is not semantically equivalent to $P$ in terms of qualified
answers. In fact, given the goal $G= p(X)$ we have $q(X) \in
\mathcal{QA'}_P(G)$, while $q(X) \not \in \mathcal{QA'}_{P'}(G).$
\end{example}

The second problem is related to multiple heads. In fact, the
unfolding that we have defined  assume that the head of a rule
matches completely with the body of another one, while in general,
during a CHR computation, a rule can match with constraints
produced by more than one rule and/or  introduced by the initial
goal. The following example illustrates this point.

\begin{example}\label{ex:unicatesta}
Let us consider the program
\[\begin{array}{rlll}
  P= \{ &r  @ p(Y)  \Leftrightarrow  q(Y), h(b).\\
  & r' @ q(Z), h(V) \Leftrightarrow  Z=V .\}\\
\end{array}
\]
where we do not consider the identifiers and the token store in
the body of rules, because we do not have propagation rules in
$P$.

The unfolding  of $r$ by using  $r'$ returns the new
rule
$$r @ p(Y) \Leftrightarrow  Y=Z, V=b, Z=V. $$ Now the
the program
\[\begin{array}{rlll}
  P'= \{ &r @ p(Y)  \Leftrightarrow Y=Z, V=b, Z=V .\\
  & r'  @  q(Z), h(V) \Leftrightarrow   Z=V .\}\\
\end{array}
\]
where we substitute the original rule by its  unfolded version is
not semantically equivalent to $P$. In fact, given the goal $G=
p(X), h(a), q(b)$, we have that  $(X=a )\in \mathcal{QA'}_P(G)$
($X=a$ is a qualified answer for $G$ in $P$) while $(X=a) \not \in
\mathcal{QA'}_{P'}(G).$
\end{example}

The final problem is related to the matching substitution. In fact,
following Definition \ref{def:unf}, there are some matching that could become
possible only at run time, and not at compile time, because a more powerful
built-in constraint store is needed. Also in this case, a rule elimination could lead to lose
possible answers as illustrated in the following example.

\begin{example}\label{ex:matching}
Let $P$ be a program
\[
\begin{array}{lclr}P&=\{&
r_1@g(X, Y) \Leftrightarrow f(X, Z)&\\
&&r_2@f(a, W) \Leftrightarrow W=b.&\\
&&r_3@f(T, J) \Leftrightarrow J=d.&\}
\end{array}
\]
where we do not consider the identifiers and the token store in the
body of rules, because we do not have propagation rules in $P$.
Let $P'$ be the program where the rule $r_1$, that is unfolded using $r_3$ in $P$,
substitutes the original $r_1$ (note that other unfolding are not possible, in
particular the rule $r_2$ can not be used to unfold $r_1$)
\[
\begin{array}{lclr}P'&=\{&
r_1@g(X, Y) \Leftrightarrow X=T, Z=J, J=d .&\\
&&r_2@f(a, W) \Leftrightarrow W=b.&\\
&&r_3@f(T, J) \Leftrightarrow J=d.&\}
\end{array}
\]
Let be $G=g(a, R)$ the goal, we can see that
$(R=b)\in\mathcal{QA'}_P(G)$ and  $(R=b)\not\in\mathcal{QA'}_{P'}(G)$ because,
with the considered goal (and consequently
the considered built-in constraint store) $r_2$ can fire in $P$ but can not fire in
$P'$.
\end{example}

We have individuated a case in which we can safely replace the
original rule $r$ by its unfolded version while maintaining the
qualified answers semantics. Intuitively, this holds when: 1)  the
constraints of the body of  $r$ can
 be rewritten only by CHR rules with a single-head, 2) there exists
no rule $v$ which has a multiple head $H$ such that a part of $H$
can match with a part of the constraints introduced in the body of
$r$ (that is, there exists no rule $v$ which can be fired by using
a part of constraints introduced in the body of $r$ plus some
other constraints) and 3) all the rules, that can be applied at run
time to the body of the original rule $r$, can also be applied at
transformation time (so unfolding avoidance for built-in constraint
store and guard-anticipation problems are solved).

Before defining formally these conditions we need some further
notations. First of all, given a rule
$r@H_1\backslash H_2 \Leftrightarrow D\,|\,\tilde A; T$, we define
two sets. The first one contains a set of pairs, whose first
component is a rule that can be used to unfold $r@H_1\backslash
H_2 \Leftrightarrow  D\,|\,\tilde A; T$, while the second one is the
sequence of the identifiers of the atoms in the body of $r$, which
are used in the unfolding.

The second set contains all the rules that can be used for the
{\em partial unfolding} of $r@H_1\backslash H_2 \Leftrightarrow
D\,|\,\tilde A; T$, namely is the set of rules that can fire by
using at least an atom in the body $\tilde A$ of the rule and
some others CHR and built-in constraints. It moreover contains
the rules that can fire if an opportune built-in constraint
store is given by the computation but that can not be unfolded
following Definition \ref{def:unf}.

\begin{definition}\label{def:Pposeneg}

Let $P$ be an annotated CHR program and let
$$
\begin{array}{rcl}
r@H_1\backslash H_2 &\Leftrightarrow&  D\,|\,\tilde A; T \mbox{ and}\\
r'@H_1'\backslash H_2' &\Leftrightarrow & D '\,|\, \tilde B; T'
\end{array}
$$
be two annotated rules, such that $r, r'\in P$ and  $r'$ is renamed
apart with respect to $r$. We define $U^{+}$ and $U^{\#}$ as follows:
\begin{enumerate}
    \item\label{uno} $(r'@H_1'\backslash H_2' \Leftrightarrow D '\,|\, \tilde B;
T', (i_1, \ldots, i_n)) \in$ $ U^{+}_P(r@H_1\backslash H_2
\Leftrightarrow  D\,|\,\tilde A; T )$
 if and only if
$r@H_1\backslash H_2 \Leftrightarrow  D\,|\,\tilde A; T $ can be
unfolded with $r'@H_1'\backslash H_2' \Leftrightarrow D '\,|\,
\tilde B; T'$ (by Definition~\ref{def:unf}) by using the sequence
of the identified atoms in $\tilde A$ with identifiers $(i_1,
\ldots, i_n)$.
    \item\label{due} $r'@H_1'\backslash H_2' \Leftrightarrow D '\,|\, \tilde B;
T'\in  U^{\#}_P(r@H_1\backslash H_2 \Leftrightarrow D\,|\,\tilde
A; T )$ if and only if one of the
following holds: \\
\begin{enumerate}
\item\label{ai} either there exist $\tilde
    A'=(\tilde A_1, \tilde A_2) \subseteq \tilde A$ and
     a built in constraint $C'$ such that
    $Fv(C') \cap Fv(r') = \emptyset$, the constraint
    $D \wedge C'$ is satisfiable,
    $CT \models (D\wedge C') \rightarrow \exists _{x}
((chr(\tilde A_1, \tilde A_2)=(H'_1,  H'_2))\wedge D')$, $r' @id
(\tilde A_1, \tilde A_2) \not \in T$ and\\
$(r'@H_1'\backslash H_2' \Leftrightarrow D '\,|\, \tilde B; T', id
(\tilde A_1, \tilde A_2)) \not  \in$ $ U^{+}_P(r@H_1\backslash H_2
\Leftrightarrow  D\,|\,\tilde A; T )$ \\
\item\label{bi} or there exist $\tilde
    A' \subseteq \tilde A$, a multiset of CHR constraints
    $H'\neq \emptyset$ and
     a built in constraint $C'$ such that $\tilde A'\neq
     \emptyset$,
    $Fv(C') \cap Fv(r') = \emptyset$, the constraint
    $D \wedge C'$ is satisfiable, $\{chr(A'), H'\}= \{K_1, K_2\}$
    and
    $CT \models (D\wedge C') \rightarrow \exists _{x}
(((K_1, K_2)=(H'_1,  H'_2))\wedge D')$.\\
\end{enumerate}
\end{enumerate}
\end{definition}
Some explanations are in order here.

The set $U^{+}$ contains all the couples composed by rules, that can be used to unfold a
fixed rule $r$, and the identifiers of the constraints considered in the unfolding,
introduced in Definition \ref{def:unf}.

Let us consider now the set $U^\#$.
The conjunction of built-in constraints $C'$
represents a generic set of built-in constraints
(said set naturally can be equal to every possible built-in constraint store that can
be generated by a real computation before the application of rule $r'$); the
condition $Fv(C')\cap Fv(r')=\emptyset$ is required to avoid free variable capture,
it represents the fresh variable rename of a rule $r'$ with respect to the computation
before the use of the $r'$ itself in an \textbf{Apply} transition;
the condition $r'@id(\tilde A_1, \tilde A_2)\not \in T$ grants the propagation rules
trivial non-termination avoidance;
the conditions $CT \models (D\wedge C') \rightarrow \exists _{x}
((chr(\tilde A_1, \tilde A_2)=(H'_1,  H'_2))\wedge D')$ and
$CT \models (D\wedge C') \rightarrow \exists _{x}
(((K_1, K_2)=(H'_1,  H'_2))\wedge D')$ secure that a strong enough built-in constraint
is possessed by the computation, before the application of rule $r'$;
the conditions $A'_1 \neq \emptyset$ and $H' \neq \emptyset$ assure respectively that
at least one constraint in the body of rule $r$ and that at least
one constraint form the initial goal or introduced by the body of other rules are unfolded;
finally the following condition
$(r'@H'_1\backslash H'_2\Leftrightarrow D'|\tilde B; T', id (\tilde A_1, \tilde A_2))\not\in
U_P^{+}(r@H_1\backslash H_2\Leftrightarrow D\mid \tilde A; T)$ is required
to avoid the consideration of the rules that can be correctly unfolded in the body of $r$.
There are two kinds of rules that are added to $U^\#$.
The first one, introduced by the Example \ref{ex:matching}, points out the
matching substitution problem (Condition \ref{ai} of Definition \ref{def:Pposeneg}).
The second kind, introduced by the Example \ref{ex:unicatesta}, points out the multiple
heads problem: the rule $r'$ can match with the body of $r$ but can also match with
other constraints introduced by the initial goal or generated by other rules
(Condition \ref{bi} of Definition \ref{def:Pposeneg}).

Note also that if $U^{+}_P(r@H_1\backslash H_2 \Leftrightarrow
D\,|\,\tilde A; T )$ contains a pair, whose first component is not
a rule with a single atom in the head, then by definition,
$U^{\#}_P(r@H_1\backslash H_2 \Leftrightarrow D\,|\,\tilde A; T )\neq
\emptyset$.

Finally, given an annotated CHR program $P$ and an annotated rule
$r@H_1\backslash H_2 \Leftrightarrow  D\,|\,\tilde A; T $, we
define \[Unf_P(r@H_1\backslash H_2 \Leftrightarrow  D\,|\,\tilde
A; T)\] as the set of all annotated rules obtained by unfolding
the rule $r@H_1\backslash H_2 \Leftrightarrow D\,|\,\tilde A; T
$ with a rule in $P$, by using Definition~\ref{def:unf}.

We can now give the central definition of this section.

\begin{definition}\textsc{(Safe rule replacement)}\label{def:nsafedel}
Let $P$ be an annotated CHR program and let $r@H_1\backslash H_2
\Leftrightarrow  D\,|\,\tilde A; T \in P$, such that the following
holds
\begin{enumerate}
\item[i)] $U^{\#}_P((r@H_1\backslash H_2 \Leftrightarrow
D\,|\,\tilde A; T  ) =\emptyset$   and

\item[ii)]  $U^{+}_P(r@H_1\backslash H_2
\Leftrightarrow D\,|\,\tilde A; T  ) \neq \emptyset$ and
\item[iii)] for each
$$r@ H_1\backslash H_2 \Leftrightarrow D'\, |\, \tilde A'; T' \in
   Unf_P (r@H_1\backslash H_2 \Leftrightarrow D\,|\,\tilde A; T)$$
we have that $CT \models D \leftrightarrow D'$.
\end{enumerate}
Then we say that the rule $r@H_1\backslash H_2 \Leftrightarrow  D\,|\,\tilde A; T $
can be safely replaced (by its unfolded version) in $P$.
\end{definition}

Some explanations are in order here.

Condition $\bf i) $ of previous definition implies that
$r@H_1\backslash H_2 \Leftrightarrow  D\,|\,\tilde A; T $ can be
safely deleted from $P$ only if:
\begin{itemize}
\item $U^{+}_P(r@H_1\backslash H_2
\Leftrightarrow D\,|\,\tilde A; T )$ contains only pairs, whose
first component is a rule with a single atom in the head.

\item  a sequence of identified atoms of
body of the rule $r$ can be used to fire a rule $r'$
only if $r$ can be unfolded with $r'$  by using the same sequence
of the identified atoms.

\end{itemize}

Condition {\bf ii)} states that exist at least one rule that unfold
the rule $r@H_1\backslash H_2 \Leftrightarrow D\,|\,\tilde A$.

Condition {\bf iii)}  states that each annotated clause obtained by the
unfolding of $r$ in $P$ must have guard equivalent to that of $r$: in fact
the condition  $CT \models D \leftrightarrow D'$ in {\bf iii)} avoids the
problems discussed in Example \ref{esempio:mau}, thus allows the anticipation
of the guard in the unfolded rule.

We can now provide the result which shows the correctness of the safe rule replacement condition.

\begin{proposition}\label{lemma:servcomplete}
Let $r@H'_1\backslash H'_2 \Leftrightarrow  D_r\,|\,\tilde K_r; T_r $ and $v$ be annotated CHR rules such that the following holds
\begin{itemize}
  \item $v$ is a rule with a single atom in the head
  \item ($r'@ H'_1\backslash H'_2 \Leftrightarrow D_{r'}, \, |\, \tilde K_{r'}; T_{r'}, i)
\in U^{+}_{\{v \}}(r@H'_1\backslash H'_2 \Leftrightarrow  D_r\,|\,\tilde K_r; T_r )$  is the result
of the unfolding of $r$ with respect to $v$, $CT \models D_r \leftrightarrow D_{r'}$ and the identified atom $\tilde k= k\#i \in \tilde K_r$.
\end{itemize}
Moreover, let $\sigma$ be a generic built-in free state such that we can
construct a derivation $\delta$ from $\sigma$ such that
\begin{itemize}
  \item $\delta$ uses at most the clauses $r$ and $v$ in the order,
  \item obtain a built-in free state $\sigma^f$ and
  \item if $v$ is used, then $v$ rewrites the atom $k\#i'$ corresponding to $k\#i \in \tilde K_r$.
\end{itemize}
Then we can use the transition \textbf{Apply'} with the clause $r'$ obtaining the state $\sigma_{r'}$ and then the built-in free state $\sigma_{r'}^f$ such that $\sigma_{r'}^f \simeq \sigma^f$.
\end{proposition}

\textsc{Proof.}
Assume that
$$
\begin{array}{rl}
\sigma&\rrarrow^ {\, r}
\sigma_r \rrarrow^{Solve^{*}}
\sigma_r^f (\rrarrow^v
\sigma_v\rrarrow^{Solve^{*}}
\sigma_v^f)\\
&\searrow_{\, r'}\sigma_{r'}\rrarrow^{Solve^{*}}
\sigma_{r'}^f\\
\end{array}
$$

The labeled arrow $\rrarrow^{Solve^{*}}$ means that only solve transitions are applied.
Moreover
\begin{itemize}
           \item if $\sigma_r^f$ has the form
$\la \tilde G, {\tt false}, T\ra$ then the derivation between the parenthesis is not present and $\sigma^f=\sigma_r^f$.
           \item the derivation between the parenthesis is present and $\sigma^f=\sigma_v^f$, otherwise.
         \end{itemize}
We have two cases since the clause $v$ is either of the form $v@k'\backslash \Leftrightarrow  D_v\,|\, \tilde P, C_v;T_v$
or of the form $v@\backslash k'\Leftrightarrow  D_v\,|\, \tilde P, C_v;T_v$. We consider only the first case. The other one is analogous and hence it is omitted.

\noindent{\textbf{Preliminaries}:} Let  $\sigma=\la (\tilde H_1,\tilde H_2,\tilde H_3) , C, T\ra_j$ be a built-in free state
and let
$r@H'_1\backslash H'_2 \Leftrightarrow  D_r\,|\,\tilde K, \tilde k, C_r; T_r $ and
$v@k'\backslash \Leftrightarrow  D_v\,|\, \tilde P, C_v;T_v$
where $\tilde k = k\#i$, $C_r$ is the conjunction of all the built-in constraints in the body
of $r$ and
\begin{equation}\label{113marzo1}
    CT \models (D_r \wedge C_r) \rightarrow chr(\tilde k)=k'\theta.
\end{equation}
Furthermore assume that $m$ is the greatest identifier which appears in the
rule $r$ and that $inst(\tilde P, T_v,m)=(\tilde P_1, T_1, m_1)$.
Then the \emph{unfolded} rule is:
\[
    r'@ H'_1\backslash H'_2
\Leftrightarrow D_r, (D_v'\theta)\, |\, \tilde K,\tilde k,\tilde
P_1, C_r, C_v, chr(\tilde k)= k'; T_{r'}
\]
where $v @id (\tilde k) \not \in T_r$,
$V\subseteq D_v$ is the greatest set of built-in constraints $c$,
such that
$CT \models (D_r\wedge C_r)\rightarrow c\theta,$
$D_v'= D_v\backslash V$, the
constraint $(D_r, (D_v'\theta))$ is satisfiable and
 then $T_{r'}=clean((\tilde K,\tilde k) ,  T_r) \cup T_1 \cup
              \{v @id (\tilde k)\}$.
 Since by hypothesis, $CT \models (D_r, (D_v'\theta))\leftrightarrow D_r$, we have that
    \begin{equation}\label{110dic2}
    CT \models (D_r\wedge C_r)\rightarrow D_v\theta \mbox{ and } D_v'\theta=\emptyset.
\end{equation}
\noindent{\textbf{The proof}:} By definition of the transition \textbf{Apply'}, we have that
\begin{equation}\label{110dic1}
    CT\models C\rightarrow \exists_x((chr(\tilde H_1,
\tilde H_2)=(H'_1, H'_2))\wedge D_r),
\end{equation}
where $ x=Fv(H'_1,H'_2)$ and
$$\begin{array}{l}
    \sigma_{r}=\la (\tilde Q_2, C_r),
          chr(\tilde H_1, \tilde H_2)= (H_1', H_2')\wedge C, T_4\ra_{j+m},
  \end{array}
  $$
     where $\tilde Q_2=(\tilde H_1,\tilde H_3,\tilde K'',\tilde k'')$,
     $((\tilde K'',\tilde k''), T_2, j+m) = inst ((\tilde K,\tilde k), T_r, j)$ and
\begin{itemize}
    \item if $H_2'=\emptyset$ then $T_4=T \cup T_2 \cup\{r @id (\tilde H_1)\}$
    \item if $H_2'\not =\emptyset$ then $T_4=T \cup T_2$.
    \end{itemize}
    Therefore, by definition
    $$ \sigma_{r}^f=\la\tilde Q_2, C_{r}^f, \, T_4\ra_{j+m}.
  $$
  where
\begin{equation}\label{13marzo21}
CT \models C_{r}^f \leftrightarrow  C_r\wedge
     chr(\tilde H_1, \tilde H_2)= (H_1', H_2')\wedge C.
\end{equation}
On the other hand, by (\ref{110dic1}), (\ref{110dic2}) and
by definition of the transition \textbf{Apply'}, we have that
$$\begin{array}{l}
    \sigma_{r'}=\la (\tilde Q, C_r, C_v,chr(\tilde k)= k'),
chr(\tilde H_1, \tilde H_2)= (H_1', H_2')\wedge C, T_3\ra_{j+m_1},
  \end{array}
  $$
     where
     $\tilde Q=(\tilde H_1,\tilde H_3,\tilde Q_1)$, with
     $inst ((\tilde K,\tilde k,\tilde P_1), T_{r'}, j)=(\tilde Q_1, T_{r'}', j+m_1)$ and
\begin{itemize}
    \item if $H_2'=\emptyset$ then $T_3=T \cup T'_{r'} \cup\{r @id (\tilde
H_1)\}$
    \item if $H_2'\not =\emptyset$ then $T_3=T \cup T'_{r'}$.
    \end{itemize}
    Therefore, by definition
    $$
    \sigma_{r'}^f=\la \tilde Q, C_{r'}^f, \, T_3\ra_{j+m_1}.$$
    where
    $$\begin{array}{l}
       CT \models C_{r'}^f \leftrightarrow  C_r\wedge
       C_v\wedge chr(\tilde k)= k'
       \wedge chr(\tilde H_1, \tilde H_2)= (H_1', H_2')\wedge C.
     \end{array}
    $$
Now, we have two possibilities
\begin{description}
  \item[($C_{r}^f = \tt false$).] In this case, by construction we have that $C_{r'}^f = \tt false$. Therefore $ \sigma_{r'}^f \simeq  \sigma_{r}^f$ and then the thesis.
  \item[($C_{r}^f    \neq \tt false$).]
  By definition, since
$chr(\tilde k)=chr(\tilde k'')$, by (\ref{113marzo1}), (\ref{110dic2}) and (\ref{110dic1}), we have that
\[\begin{array}{ll}
    CT\models  & (C_r\wedge
     chr(\tilde H_1, \tilde H_2)= (H_1', H_2')\wedge C)
     \rightarrow
     \exists_y((chr(\tilde k'')=k')\wedge D_v),
  \end{array}
 \]
where $ y=Fv(k)$. Therefore
by (\ref{13marzo21})
\[CT\models  C_{r}^f
     \rightarrow  (\exists_y((chr(\tilde k'')=k')\wedge D_v))
 \]

and since by hypothesis $v$ rewrites the atom $\tilde k''$ corresponding to $\tilde k \in \tilde K_r$, we have that
$$\begin{array}{ll}
    \sigma_{v}=&\la (Q_3 ,  C_v),
     chr(\tilde k'')= k'\wedge C_r\wedge
     chr(\tilde H_1, \tilde H_2)= (H_1', H_2')\wedge C, \, T_5\ra_{m_1},
  \end{array}
  $$
     where
     $\tilde Q_3=(\tilde H_1,\tilde H_3,\tilde K'',\tilde k'', \tilde P_2)$, with
     $inst (\tilde P, T_v, j+m) = (\tilde P_2, T_v', m_1)$ and $T_5=T_4 \cup T_v' \cup\{v @id (\tilde
k'')\}$.

    Finally by definition, we have that
$$
    \sigma_{v}^f=\la \tilde Q_3,C_{v}^f, \, T_5\ra_{m_1},
$$
  where
  $$\begin{array}{l}
    CT \models C_{v}^f \leftrightarrow C_v \wedge
     chr(\tilde k'')= k'\wedge C_r\wedge
     chr(\tilde H_1, \tilde H_2)= (H_1', H_2')\wedge C.
  \end{array}
  $$

  If $C_{v}^f =\tt false $ then the proof is analogous to the previous case and hence it is omitted.

  Otherwise, the proof is analogous to that given for Proposition \ref{prop:servequality} and hence it is omitted.
\end{description}
\noindent{$\Box$}

\begin{theorem}\label{lemma:n1completeness}
Let $P$ be an annotated program, \\ $r@H_1\backslash H_2
\Leftrightarrow D\,|\,\tilde A; T$ be a rule in $P$ such that
$r@H_1\backslash H_2 \Leftrightarrow  D\,|\,\tilde A; T $ can be safely replaced
in $P$ according to Definition \ref{def:nsafedel}. Assume also that
\[\begin{array}{l}
  P'  =  (P\setminus   \{(r@H_1\backslash H_2 \Leftrightarrow D\,|\,\tilde A; T)\} ) \, \cup
   Unf_{P}(r@H_1\backslash H_2 \Leftrightarrow D\,|\,\tilde A; T).
\end{array}
\]

Then $\mathcal{QA'}_{P}(G)=\mathcal{QA'}_{P'}(G)$ for any arbitrary
goal $G$.
\end{theorem}

\textsc{Proof.} By using a straightforward inductive argument and by
Proposition \ref{lemma:equality}, we have that
$\mathcal{QA'}_{P}(G)=\mathcal{QA'}_{P''}(G)$ where
 \[
  P'' = P \, \cup \, Unf_{P}(r@H_1\backslash H_2 \Leftrightarrow D\,|\,\tilde A; T),
\]
for any arbitrary goal $G$.

Then to prove the thesis, we have only to prove that
\[\mathcal{QA'}_{P'}(G) = \mathcal{QA'}_{P''}(G).\]
We prove the two inclusions separately.
\begin{description}
  \item[{\bf ($\mathcal{QA'}_{P'}(G) \subseteq \mathcal{QA'}_{P''}(G)$) }]
  The proof is by contradiction. Assume that there exists $(K'\wedge d') \in \mathcal{QA'}_{P'}(G) \setminus \mathcal{QA'}_{P''}(G)$. By definition there exists a derivation
  \[\delta=\langle I_0^m (G),
{\tt true}, \emptyset\rangle_m\rightarrow^{*}_{\omega'_t}
\langle \tilde K, d, T\rangle_n\not\rightarrow_{\omega'_t}\]
in $P'$, such that $(K'\wedge d') =
\exists _{-Fv(G)}(chr( \tilde K)\wedge d)$. Since $P' \subseteq P''$, we have that there exists the derivation
\[\langle I_0^m (G),
{\tt true}, \emptyset\rangle_m\rightarrow^{*}_{\omega'_t}
\langle \tilde K, d, T\rangle_n\] in $P''$. Moreover, since $P'' =P '\cup \{(r@H_1\backslash H_2 \Leftrightarrow D\,|\,\tilde A; T)\} $ and $(K'\wedge d')\not  \in \mathcal{QA'}_{P'}(G)$, we have that there exists a derivation step $\langle \tilde K, d, T\rangle_n\rightarrow_{\omega'_t}
\langle \tilde K_1, d_1, T_1\rangle_{n_1}$ by using the clause $r@H_1\backslash H_2 \Leftrightarrow D\,|\,\tilde A; T$. \\
Since $r@H_1\backslash H_2 \Leftrightarrow  D\,|\,\tilde A; T $ can be safely replaced
in $P$, we have that there exists \[r@H_1\backslash H_2 \Leftrightarrow  D'\,|\,\tilde A'; T' \in Unf_{P}(r@H_1\backslash H_2 \Leftrightarrow D\,|\,\tilde A; T)\] such that
$CT \models D \leftrightarrow D'$. \\
Then there exists a derivation step $\langle \tilde K, d, T\rangle_n\rightarrow_{\omega'_t}
\langle \tilde K_2, d_2, T_2\rangle_{n_2}$ in $P'$ (by using the clause $r@H_1\backslash H_2 \Leftrightarrow  D'\,|\,\tilde A'; T' \in P'$) and then we have a contradiction.

  \item[{\bf ($\mathcal{QA'}_{P''}(G) \subseteq \mathcal{QA'}_{P'}(G)$) }]
  First of all, observe that by Proposition \ref{prop:solonorm},
  $\mathcal{QA'}(P'')$ can be calculated by
considering only normal terminating derivations. \\
Moreover, since  by hypothesis
$r@H_1\backslash H_2 \Leftrightarrow D\,|\,\tilde A; T$ can be safely replaced
in $P$, following Definition \ref{def:nsafedel} (Safe rule replacement), we have that
\[Unf_{P}(r@H_1\backslash H_2 \Leftrightarrow D\,|\,\tilde A; T) \neq \emptyset\]
and
\[ r@ H_1\backslash H_2
\Leftrightarrow D'\, | \,\tilde B; T' \in
     Unf_{P}(r@H_1\backslash H_2 \Leftrightarrow D\,|\,\tilde A; T)\] if and only if there exists a rule
     $v\in P$ with a single atom in the head such that
 \[ (r@ H_1\backslash H_2 \Leftrightarrow D' \, |\, \tilde B; T', i)
\in U^{+}_{\{v \}}(r@H_1\backslash H_2 \Leftrightarrow D\,|\,\tilde A; T ),\]
and $CT \models D \leftrightarrow D'$.\\
Then for each normal terminating derivation $\delta$, which uses the clause $r@H_1\backslash H_2 \Leftrightarrow D\,|\,\tilde A; T$ after the application of $r@H_1\backslash H_2 \Leftrightarrow D\,|\,\tilde A; T$, we obtain the state $\sigma_{r}$ and then the built-in free state $\sigma_{r}^f=\la \tilde K, C , T''\ra _{m}$. Now, we have two cases

\begin{itemize}
\item either $CT \models C \leftrightarrow \tt false$
\item or $CT \models C \not \leftrightarrow \tt false$. In this case, since by hypothesis
$r@H_1\backslash H_2 \Leftrightarrow D\,|\,\tilde A; T$ can be safely replaced
in $P$, following Definition \ref{def:nsafedel}, we have there exists an atom
$\tilde k \in \tilde A$, such that $\tilde k$ is rewritten in $\delta$ by using a clause $v \in P$,
$(r@ H_1\backslash H_2 \Leftrightarrow D' \, |\, \tilde B; T', id(\tilde k))
\in U^{+}_{\{v \}}(r@H_1\backslash H_2 \Leftrightarrow D\,|\,\tilde A; T )$ and $CT \models D \leftrightarrow D'$.
Without loss of generality we can assume that in the derivation $\delta$, the clause $v$ is applied to the considered state $\sigma_{r}^f=\la \tilde K, C , T''\ra _{m}$ (in order to rewrite the atom $\tilde k'$ corresponding to $\tilde k\in \tilde A$).

\end{itemize}
In both the cases, the proof is straightforward, by using previous observations and by Proposition \ref{lemma:servcomplete}.
\end{description}

\noindent{$\Box$}

Of course, previous result can be applied to a sequence of program transformations.
Let us define such a sequence as follows.

\begin{definition}[U-sequence]\label{def:uno}
Let $P$ be an annotated CHR program. An \emph{U-sequence} of programs
starting from $P$ is a sequence of annotated CHR programs $P_0, \ldots,
P_n$, such that
\[
\begin{array}{lll}
  P_0 & = & P  \mbox{ and }\\
  P_{i+1}& = &  P_i \setminus   \{(r@H_1\backslash H_2 \Leftrightarrow D\,|\,\tilde A; T)\} ) \, \cup\\
  &  & Unf_{P_i}(r@H_1\backslash H_2 \Leftrightarrow D\,|\,\tilde A; T), \\
\end{array}
 \]
 where $i \in [0,n-1]$, $(r@H_1\backslash H_2 \Leftrightarrow D\,|\,\tilde A;
 T)\in P_i$ and $(r@H_1\backslash H_2 \Leftrightarrow D\,|\,\tilde A;
 T)$  is
safety deleting from $P_i$
\end{definition}

Then from  Theorem~\ref{lemma:n1completeness} and
Proposition~\ref{lemma:nequality} we have immediately the
following.

\begin{corollary}\label{lemma:ncompleteness}
Let $P$ be a program and let $P_0, \ldots, P_n$ be an U-sequence starting from $Ann(P)$. Then
$\mathcal{QA}_{P}(G)=\mathcal{QA'}_{P_n}(G)$ for any arbitrary goal
$G$.
\end{corollary}

{\sc Proof.} Proposition~\ref{lemma:nequality} proves that $\mathcal{QA}_{P}(G)=\mathcal{QA}'_{P_0}(G)$, for every goal $G$, where  $P_0=Ann(P)$.
Theorem \ref{lemma:n1completeness} proves that, for every goal $G$ and for $i \in [1, n-1]$,
$\mathcal{QA}'_{P_i}(G)=\mathcal{QA}'_{P_{i+1}}(G)$. Then the proof follows by a straightforward inductive argument.
\\
\noindent{$\Box$}

\section{Confluence and Termination}\label{sec:confluence&termination}

It is also possible to prove that our unfolding preserves normal termination and normal confluence.

The formal definition of termination from \cite{Fru04}
is introduced and adapted to our $\omega_t'$ semantics.

\begin{definition}[Termination]
A CHR program $P$ is called \emph{terminating}, if there are no infinite computations.
\end{definition}

\begin{definition}[Normal Termination]
A (possibly annotated) CHR program $P$ is called  \emph{normal terminating}, if there are no infinite normal computations.
\end{definition}

\begin{proposition}[Normal Termination]\label{prop:termination} Let $P$ be a CHR program and
let $P_0, \ldots, P_n$ be an U-sequence starting from $Ann(P)$. $P$ satisfies
normal termination if and only if $P_n$  satisfies normal termination.
\end{proposition}
\textsc{Proof.}
By Lemma \ref{lemma:intermequiv}, we have that $P$ is normal terminating if and only if $Ann(P)$ is normal terminating.
Moreover from Proposition \ref{prop:servequality} and Proposition \ref{lemma:servcomplete} and by using a straightforward inductive argument, we have that
for each $i=0, \ldots,n-1$, if $P_i$ satisfies normal termination if and only if $P_{i+1}$ satisfies the normal termination too and then the thesis.
\\
\noindent{$\Box$} \\

When (standard) termination is considered instead of normal termination, program
transformation, defined in Definition \ref{def:uno} (U-sequence), can introduce
problems connected to the guard elimination process of Definition \ref{def:unf}
(Unfold) as showed in the following example.

\begin{example}
Let us consider the following program:
\[
\begin{array}{lcll}
P&=\{&r_1@p(X)\Leftrightarrow \mid X=a, q(X).&\\
&&r_2@q(Y)\Leftrightarrow Y=a \mid r(Y).&\\
&&r_3@r(Z)\Leftrightarrow Z=d \mid p(Z).&\}
\end{array}
\]
where we do not consider the identifiers and the token store in the
body of rules, because we do not have propagation rules in $P$.
Then the following possible unfolded program $P'$, where the previous
$r_1$ is unfolded using $r_2$ (following Definition \ref{def:unf}) and where the
(original clause) $r_1 \in P$ is deleted because safe rule replacement holds, so results of Theorem
\ref{lemma:n1completeness} can be applied, is given:
\[
\begin{array}{lcll}
P'&=\{&r_1@p(X)\Leftrightarrow \mid X=a, X=Y, r(Y).&\\
&&r_2@q(Y)\Leftrightarrow Y=a \mid r(Y).&\\
&&r_3@r(Z)\Leftrightarrow Z=d \mid p(Z).&\}
\end{array}
\]
It is easy to check that the program $P$ satisfies the (standard) termination.
If instead the program $P'$ and the start goal $(V=d,p(V))$ are
considered, the following state can be reached
$$\la  (X=a, p(Z)\#3),(V=d, V=X, X= Y, Y=Z), \emptyset \ra_4$$
where $r_1,r_3$ (in the order) can be applied infinite times if the built-in constraint $X=a$
is not moved by \textbf{Solve'} rule into the built-in store, where it would be
evaluated. This can happen because of the non determinism in rule application of
$\omega'_t$ semantics.
\end{example}

The confluence property guarantees that any computation for a goal
results in the same final state, no matter which of the applicable
rules are applied \cite{AF04}. This means that $\mathcal{QA}_P(G)$ has
cardinality  at the most one for each goal $G$.
The formal definition of confluence from \cite{Fru04}
is introduced and adapted to our $\omega_t'$ semantics.
Confluence is considered only for normal terminating programs and in this case $\mathcal{QA}_P(G)$ has
cardinality exactly one for each goal $G$.
In the following $\mapsto^{*}$ means either $\rrarrow_{\omega_t} $ or $\rrarrow_{\omega'_t}$.

\begin{definition}[Confluence]
A CHR [annotated] program is \emph{confluent} if for all states
$\sigma, \sigma_1, \sigma_2$:
if $\sigma\mapsto^{*} \sigma_1$ and $\sigma\mapsto^{*} \sigma_2$ then exist states $\sigma_f'$
and $\sigma_f''$ such that $\sigma_1 \mapsto^{*}\sigma_f'$ and $\sigma_2\mapsto^{*}\sigma_f''$ and
$\sigma_f'$ and $\sigma_f''$ are identical up to renaming of local variables,
identifiers and logical equivalence of built-in constraints.
\end{definition}

We now introduce the concept of normal confluence.

\begin{definition}
Let
$\sigma_1, \sigma_2 \in {\it Conf_t} ({\it Conf'_t})$ and let $V$ be a set of variables. $\sigma_1\simeq\,'_{V}\,\sigma_2$ if the following holds:
\begin{itemize}
  \item either $\sigma_1$ and $\sigma_2$ are both failed configurations
  \item or $\sigma_1$ and $\sigma_2$ are identical up to renaming of variables not in $V$,
identifiers, up to cleaning the token store (namely, up to deleting from the token store all the tokens for
which at least one identifier is not present in the set of
identified CHR constraints) and logical equivalence of built-in constraints.
\end{itemize}
\end{definition}

\begin{definition}[Normal Confluence]
A CHR [annotated] program is \emph{normal confluent} if for all states
$\sigma, \sigma_1, \sigma_2$:
if there exist two normal derivations $\sigma\mapsto^{*} \sigma_1$ and $\sigma\mapsto^{*} \sigma_2$ then $\sigma_1 \mapsto^{*}\sigma_f'$ and $\sigma_2\mapsto^{*}\sigma_f''$, where $\sigma_f'\simeq\,'_{Fv(\sigma)}\,\sigma_f''$.
\end{definition}

Observe that, by definition, if a CHR [annotated] program is confluent, then it is normal confluent.

\begin{lemma}\label{relrel}
Let $\sigma, \sigma'$ be final configurations in ${\it Conf_t}$, $\sigma_1, \sigma_2, \sigma'_1, \sigma'_2 \in {\it Conf'_t}$ and let $V$ be a set of variables.
\begin{itemize}
  \item If $\sigma_1\equiv \sigma$, $\sigma'_1\equiv \sigma'$ then  $\sigma_1\simeq\,'_{V}\sigma'_1$ if and only if $\sigma\simeq\,'_{V}\,\sigma'$.
  \item If $\sigma_1\simeq \sigma_2$, $\sigma'_1\simeq \sigma'_2$ and  $\sigma_1\simeq\,'_{V}\,\sigma'_1$ then $\sigma_2\simeq\,'_{V}\,\sigma'_2$.
\end{itemize}
\end{lemma}
\textsc{Proof.}
The proof of the first statement follows by definition of $\equiv$ and by observing that if $\sigma$ is a final configuration in ${\it Conf_t}$, then $\sigma$
has either the form $\langle  G,
\tilde S, {\tt false} , T\rangle_n$  or it
has the form $\langle \emptyset,\tilde S, c,T \rangle_n$.

The proof of the second statement is straightforward, by observing that if $\sigma_1\simeq \sigma_2$, then
$\sigma_1 \simeq\,'_{V}\,\sigma_2$ for each set of variables $V$.

\begin{lemma}\label{conflnorm}
Let $P$ be a CHR [annotated] program. $P$ is normal confluent if for all states
$\sigma, \sigma_1, \sigma_2$:
if there exist two normal derivations $\sigma\mapsto^{*} \sigma_1$ and $\sigma\mapsto^{*} \sigma_2$ then there exists two normal derivations $\sigma_1 \mapsto^{*}\sigma_f'$ and $\sigma_2\mapsto^{*}\sigma_f''$ such that $\sigma_f'\simeq\,'_{Fv(\sigma)}\,\sigma_f''$.
\end{lemma}
\textsc{Proof.}
 In the following we assume that $P$ is a CHR annotated program.
If $P$ is a standard CHR program, the proof is analogous and hence it is omitted.

The proof is by contradiction.
Assume that $P$ is normal confluent and there exists the states
$\sigma, \sigma_1, \sigma_2$ such that there exists  two normal derivations $\sigma\mapsto^{*} \sigma_1$ and $\sigma\mapsto^{*} \sigma_2$ such that there are no two normal derivations $\sigma_1 \mapsto^{*}\sigma_f'$ and $\sigma_2\mapsto^{*}\sigma_f''$ such that
$\sigma_f'\simeq\,'_{Fv(\sigma)}\,\sigma_f''$.
Since $P$ is normal confluent, there exists two built-in free states $\sigma'_1$ and $\sigma'_2$ such that $\sigma_1 \mapsto^{*}\sigma'_1$ and $\sigma_2\mapsto^{*}\sigma'_2$ and
$\sigma'_1 \simeq\,'_{Fv(\sigma)} \,\sigma'_2$.

Let $\sigma^f_1=\la\tilde K_1 , D_1, T_1\ra_{o_1}$ ($\sigma^f_2=\la\tilde K_2 , D_2, T_2\ra_{o_2}$) be the built-in free state obtained from $\sigma'_1$ ($\sigma'_2$) by evaluating all the built-in constraints in $\sigma'_1$ ($\sigma'_2$). Since $\sigma'_1\simeq\,'_{Fv(\sigma)} \,\sigma'_2$ it is easy to check that $\sigma^f_1\simeq\,'_{Fv(\sigma)} \sigma^f_2$.

Now, we have two possibilities
\begin{itemize}
  \item $D_1 \neq \tt false$ and $D_2 \neq \tt false$. In this case, it is easy to check that there exists two normal derivation $\sigma_1 \mapsto^{*}\sigma^f_1$ and $\sigma_2\mapsto^{*}\sigma^f_2$ obtained form $\sigma_1 \mapsto^{*}\sigma'_1$ and $\sigma_2\mapsto^{*}\sigma'_2$ by evaluating the built-in constraints as soon as possible.
  \item $D_1 = \tt false$ and $D_2 = \tt false$. In this case, we that there exists two normal derivations, such that $\sigma_1 \mapsto^{*}\sigma''_1\not \mapsto$ and
      $\sigma_2\mapsto^{*}\sigma''_2\not \mapsto$,
      where
      \[\sigma''_1=\la \tilde K'_1 , {\tt false} , T'_1 \ra_{o'_1} \mbox{ and }
      \sigma''_2 = \la \tilde K'_2 , \tt false , T'_2 \ra_{o'_2}.\]
 \end{itemize}
In both the case, by definition of $\simeq\,'_{Fv(\sigma)}$, we have a contradiction to the hypothesis that there are no two normal derivations $\sigma_1 \mapsto^{*}\sigma_f'$ and $\sigma_2\mapsto^{*}\sigma_f''$ such that
$\sigma_f'\simeq\,'_{Fv(\sigma)}\sigma_f''$ and then the thesis.

\noindent{$\Box$}

The following Lemma is a straightforward consequence of the previous one.

\begin{lemma}\label{conflnormterm}
Let $P$ be a CHR [annotated] normal terminating program. If $P$ is not normal confluent there exist a state
$\sigma$ and two normal derivations $\sigma\mapsto^{*} \sigma_1^f\not \mapsto^{*} $ and $\sigma\mapsto^{*} \sigma_2^f\not\mapsto^{*} $ such that
$\sigma_1^f \not \simeq\,'_{Fv(\sigma)}\sigma_2^f$.
\end{lemma}
\textsc{Proof.}
Assume that $P$ is not normal confluent. By Lemma \ref{conflnorm},
there exist the states
$\sigma, \sigma_1, \sigma_2$ such that
there exist two normal derivations $\sigma\mapsto^{*} \sigma_1$ and $\sigma\mapsto^{*} \sigma_2$ and there are no two normal derivations $\sigma_1 \mapsto^{*}\sigma'_1$ and $\sigma_2\mapsto^{*}\sigma'_2$ in $P$ such that $\sigma_1'\simeq\,'_{Fv(\sigma)}\sigma_2'$.

Since $P$ is normal terminating, there are two normal derivations
\[\sigma \mapsto^{*}\sigma_1 \mapsto^{*}\sigma_1^f \not \mapsto^{*} \mbox{ and } \sigma\mapsto^{*}\sigma_2\mapsto^{*}\sigma_2^f\not\mapsto^{*}
\]
in $P$. By previous observation, we have that $\sigma_1^f \not \simeq\,'_{Fv(\sigma)} \sigma_2^f$ and then the thesis.

\noindent{$\Box$}

\begin{corollary}[Normal Confluence] Let $P$ be a normal terminating CHR program and let $P_0, \ldots, P_n$
be an U-sequence starting from $Ann(P)$. $P$ satisfies normal confluence if and only if $P_n$
satisfies normal confluence too.
\end{corollary}
\textsc{Proof.}
\begin{itemize}
  \item Assume that $P$ is a normal terminating CHR program and that $P$ satisfies normal confluence. We prove that $P_n$
satisfies normal confluence too.
First of all, observe, that by hypothesis and by Proposition \ref{prop:termination}, we have that $P_n$ is normal terminating.

Let us assume by contrary that $P_n$ does not satisfy normal confluence.
By Lemma \ref{conflnormterm},  there exists a state $\sigma = \langle (\tilde K,D),C,
T\rangle_o$ and two normal derivations $\sigma\rrarrow_{\omega'_t}^{*} \sigma_1^f \not \rrarrow_{\omega'_t} $ and $\sigma\rrarrow_{\omega'_t}^{*}\sigma_2^f\not \rrarrow_{\omega'_t}$ in $P_n$ such that
$\sigma_1^f\not \simeq\,'_{Fv(\sigma)} \sigma_2^f$.

Then, by using arguments similar to that given in Proposition \ref{prop:servequality}, we have that there exist two normal derivations
\[\sigma \rrarrow_{\omega'_t}^{*}\sigma'_1\not \rrarrow_{\omega'_t} \mbox{ and } \sigma \rrarrow_{\omega'_t}^{*}\sigma'_2\not \rrarrow_{\omega'_t}\]
in $P_0$, where
$\sigma'_1 \simeq \sigma_1^f$ and $\sigma'_2 \simeq \sigma_2^f$.

Therefore,by  Proposition~\ref{lemma:nequality}  there exist two normal derivations
\[\sigma'\rrarrow_{\omega_t}^{*}\sigma'_f\not \rrarrow_{\omega_t} \mbox{ and } \sigma'\rrarrow_{\omega_t}^{*}\sigma''_f\not \rrarrow_{\omega_t}\]
in $P$, where
$\sigma'=\langle D,\tilde  K,C, T\rangle_{o+1}$, $\sigma_f' \equiv \sigma'_1$ and $\sigma_f'' \equiv \sigma'_2$. Since  by hypothesis $P$ is normal confluent, we have that $\sigma'_f\simeq\,'_{Fv(\sigma')} \sigma''_f$ and therefore, by Lemma \ref{relrel} we have a contradiction to the fact that $\sigma_1^f\not \simeq\,'_{Fv(\sigma)} \sigma_2^f$ and then the thesis.

  \item Assume that $P$ is a normal terminating CHR program and that $P_n$ satisfies normal confluence. The proof that $P$
satisfies normal confluence is analogous to the previous one, by using Proposition \ref{lemma:servcomplete} instead of Proposition \ref{prop:servequality}.
\end{itemize}

\noindent{$\Box$}

\section{Weak safe rule replacement}
In this subsection we consider only normal terminating and normal confluent programs and we give a weaker condition in order to safely replace the original rule $r$ by its unfolded version while maintaining the qualified answers semantics. Intuitively this holds when there exists a rule obtained by the unfolding of $r$ in $P$ whose guard is equivalent to that of $r$.

\begin{definition}\textsc{(Weak safe rule replacement)}\label{def:wsafedel}
Let $P$ be an annotated CHR program and let $r@H_1\backslash H_2
\Leftrightarrow  D\,|\,\tilde A; T \in P$, such that there exists
$$r@ H_1\backslash H_2 \Leftrightarrow D'\, |\, \tilde A'; T' \in
   Unf_P (r@H_1\backslash H_2 \Leftrightarrow D\,|\,\tilde A; T)$$
with $CT \models D \leftrightarrow D'$.

Then we say that the rule $r@H_1\backslash H_2 \Leftrightarrow  D\,|\,\tilde A; T $
can be weakly safe replaced (by its unfolded version) in $P$.
\end{definition}

\begin{definition}[WU-sequence]\label{def:wuno}
Let $P$ be an annotated CHR program. An \emph{WU-sequence} of programs
starting from $P$ is a sequence of annotated CHR programs $P_0, \ldots,
P_n$, such that
\[
\begin{array}{lll}
  P_0 & = & P  \mbox{ and }\\
  P_{i+1}& = &  P_i \setminus   \{(r@H_1\backslash H_2 \Leftrightarrow D\,|\,\tilde A; T)\} ) \, \cup\\
  &  & Unf_{P_i}(r@H_1\backslash H_2 \Leftrightarrow D\,|\,\tilde A; T), \\
\end{array}
 \]
 where $i \in [0,n-1]$, $(r@H_1\backslash H_2 \Leftrightarrow D\,|\,\tilde A;
 T)\in P_i$ and $(r@H_1\backslash H_2 \Leftrightarrow D\,|\,\tilde A;
 T)$  is
weakly safety deleting from $P_i$.
\end{definition}

\begin{proposition}\label{prop:wterm}
Let $P$ be an annotated CHR program and let $r@H_1\backslash H_2 \Leftrightarrow  D\,|\,\tilde A; T \in P$ such that $r@H_1\backslash H_2 \Leftrightarrow  D\,|\,\tilde A; T $
can be weakly safe replaced (by its unfolded version) in $P$. Moreover let
\[P'  =  (P\setminus   \{(r@H_1\backslash H_2 \Leftrightarrow D\,|\,\tilde A; T)\} ) \, \cup
   Unf_{P}(r@H_1\backslash H_2 \Leftrightarrow D\,|\,\tilde A; T)\]

    If $P$ is normal terminating and normal confluent then  $P'$ is normal terminating and normal confluent too.
\end{proposition}
\textsc{Proof.}
First, we prove that if $P$ is normal terminating and normal confluent then  $P''$ is normal terminating and normal confluent too, where
\[P''= P \, \cup
   Unf_{P}(r@H_1\backslash H_2 \Leftrightarrow D\,|\,\tilde A; T). \]
   Then we prove that if $P''$ is normal terminating and
   normal confluent then $P'$ is normal terminating and normal confluent and then the thesis.
\begin{itemize}
   \item Assume that $P$ is normal terminating.
  The proof of the normal termination of $P''$ follows by Proposition \ref{prop:servequality}.

   \item Now, assume that $P$ is normal terminating and normal confluent and by the contrary that $P''$ does not satisfy normal confluence.

       By Lemma \ref{conflnormterm} and since by previous result $P''$ is normal terminating, there exist a state
$\sigma$ and two normal derivations
\[\sigma\rrarrow_{\omega'_t}^{*}\sigma'_f \not \rrarrow_{\omega'_t} \mbox{ and } \sigma\rrarrow_{\omega'_t}^{*}\sigma''_f\not \rrarrow_{\omega'_t}
\] in $P''$
such that
$\sigma'_f \not \simeq\,'_{Fv(\sigma)}\sigma''_f$.

Then, by using arguments similar to that given in Proposition \ref{prop:servequality} and since $P\subseteq P''$, we have that there exist two normal derivations
\[\sigma\rrarrow_{\omega'_t}^{*}\sigma_1^f \not \rrarrow_{\omega'_t}\mbox{ and } \sigma'\rrarrow_{\omega'_t}^{*}\sigma_2^f
\not \rrarrow_{\omega'_t}\]
in $P$, where
 $\sigma'_f \simeq \sigma_1^f$ and $\sigma_f'' \simeq \sigma_2^f$. Since by hypothesis $P$ is normal confluent, we have that $\sigma_1^f \simeq\,'_{Fv(\sigma)}\sigma_2^f$.
  Therefore, by Lemma \ref{relrel} we have a contradiction to the assumption that there exist two states $\sigma'_f$ and $\sigma''_f$ as previously defined.
 \end{itemize}

 \noindent Now, we prove that if $P''$ is normal terminating and normal confluent then $P'$ is normal terminating and normal confluent too and then the thesis.
 \begin{itemize}
   \item If $P''$ is normal terminating then, since $P'\subseteq P''$, we have that
  $P'$ is normal terminating too.
   \item Now, assume that $P''$ is normal terminating and normal confluent and  by the contrary that $P'$ does not satisfy normal confluence. By Lemma \ref{conflnormterm} and since by previous result $P'$ is normal terminating, there exist a state
$\sigma$ and two normal derivations
\[\sigma\rrarrow_{\omega'_t}^{*}\sigma_1^f\not \rrarrow_{\omega'_t} \mbox{ and } \sigma\rrarrow_{\omega'_t}^{*}\sigma_2^f\not \rrarrow_{\omega'_t}
\]
 in $P'$ such that
$\sigma_1^f \not \simeq\,'_{Fv(\sigma)}\sigma_2^f$.

Since $P' \subseteq P''$, we have that there exist two normal derivations
\[\sigma\rrarrow_{\omega_t}^{*}\sigma_1^f \mbox{ and } \sigma'\rrarrow_{\omega_t}^{*}\sigma_2^f\]
in $P''$. Then, since $P''$ is normal confluent and $P''= P '\cup \{r@H_1\backslash H_2 \Leftrightarrow D\,|\,\tilde A; T\}$ there exists $i \in [1,2]$ such that
$\sigma_i^f\rrarrow_{\omega'_t}\sigma'$ in $P''$ by using the clause $r@H_1\backslash H_2 \Leftrightarrow D\,|\,\tilde A; T \in (P'' \setminus P')$. In this case, by definition of weakly safe replacement, there exists
$$r@ H_1\backslash H_2 \Leftrightarrow D'\, |\, \tilde A'; T' \in
   Unf_P (r@H_1\backslash H_2 \Leftrightarrow D\,|\,\tilde A; T)$$
with $CT \models D \leftrightarrow D'$. Therefore $\sigma_i^f\rrarrow_{\omega'_t}\sigma''$ in $P'$ by using the clause $r@ H_1\backslash H_2 \Leftrightarrow D'\, |\, \tilde A'; T'$ and then we have a contradiction.
 \end{itemize}

\noindent{$\Box$}

\begin{theorem}\label{prop:wqualified}
Let $P$ be a normal terminating and normal confluent annotated program and let $r@H_1\backslash H_2
\Leftrightarrow D\,|\,\tilde A; T$ be a rule in $P$ such that
$r@H_1\backslash H_2 \Leftrightarrow  D\,|\,\tilde A; T $ can be weakly safely replaced
in $P$ according to Definition \ref{def:wsafedel}. Assume also that
\[\begin{array}{l}
  P'  =  (P\setminus   \{(r@H_1\backslash H_2 \Leftrightarrow D\,|\,\tilde A; T)\} ) \, \cup
   Unf_{P}(r@H_1\backslash H_2 \Leftrightarrow D\,|\,\tilde A; T).
\end{array}
\]

Then $\mathcal{QA'}_{P}(G)=\mathcal{QA'}_{P'}(G)$ for any arbitrary
goal $G$.
\end{theorem}

\textsc{Proof.} Analogously to Theorem \ref{lemma:n1completeness}, we can prove that
$\mathcal{QA'}_{P}(G)=\mathcal{QA'}_{P''}(G)$ where
 \[
  P'' = P \, \cup \, Unf_{P}(r@H_1\backslash H_2 \Leftrightarrow D\,|\,\tilde A; T),
\]
for any arbitrary goal $G$.

Then to prove the thesis, we have only to prove that
\[\mathcal{QA'}_{P'}(G) = \mathcal{QA'}_{P''}(G).\]
We prove the two inclusions separately.
\begin{description}
  \item[{\bf ($\mathcal{QA'}_{P'}(G) \subseteq \mathcal{QA'}_{P''}(G)$) }]
  The proof is the same of the case $\mathcal{QA'}_{P'}(G) \subseteq \mathcal{QA'}_{P''}(G)$ of Theorem \ref{lemma:n1completeness} and hence it is omitted.

  \item[{\bf ($\mathcal{QA'}_{P''}(G) \subseteq \mathcal{QA'}_{P'}(G)$) }]
  The proof is by contradiction. Assume that there exists $(K'\wedge d') \in \mathcal{QA'}_{P''}(G) \setminus \mathcal{QA'}_{P'}(G)$. Since $P''$ is normal terminating and normal confluent and since by previous point
  $\mathcal{QA'}_{P'}(G) \subseteq \mathcal{QA'}_{P''}(G)$, we have that $\mathcal{QA'}_{P'}(G)= \emptyset$.
  This means that each normal derivation in $P'$ is not terminating and hence, by using Proposition \ref{prop:wterm}, we have a contradiction.

\end{description}
\noindent{$\Box$}

\begin{corollary}\label{lemma:wcompleteness}
 Let $P$ be a normal terminating and normal confluent program and let $P_0, \ldots, P_n$ be an WU-sequence starting from                                                 $Ann(P)$. Then
$\mathcal{QA}_{P}(G)=\mathcal{QA'}_{P_n}(G)$ for any arbitrary goal
$G$.
\end{corollary}

{\sc Proof.} We prove by induction on $i$, that for each $i \in [1,n]$, $P_i$ is a normal terminating and normal confluent program and that $\mathcal{QA}_{P}(G)=\mathcal{QA'}_{P_n}(G)$ for any arbitrary goal
$G$.
\begin{description}
  \item[$i=0$)] Proposition~\ref{lemma:nequality} proves that the qualified answer for
a program $P$ and its annotated version $P_0=Ann(P)$, fixed a start goal, is the same. Moreover, by using Proposition~\ref{lemma:nequality} it is easy to check that if $P$ is normal terminating and normal confluent, then $P_0$ is normal terminating and normal confluent.
  \item[$i>0$)] Assume that the thesis holds for $i-1$, namely $P_{i-1}$ is a normal terminating and normal confluent program and that $\mathcal{QA}_{P}(G)=\mathcal{QA'}_{P_{i-1}}(G)$ for any arbitrary goal
$G$. Then, by using Proposition \ref{prop:wterm}, we have that $P_{i}$ is a normal terminating and normal confluent program. Moreover by Theorem \ref{prop:wqualified} we have that $\mathcal{QA'}_{P_i}(G)=\mathcal{QA'}_{P_{i-1}}(G)$ Therefore by inductive hypothesis $\mathcal{QA}_{P}(G)=\mathcal{QA'}_{P_{i-1}}(G)$ and then the thesis.
\end{description}

\section{Conclusions}\label{sec:conclusion_and_future}

In this paper we have defined an unfold operation for CHR which preserves the qualified answers of a program.

This was obtained by transforming a CHR program into an
annotated one which is then unfolded. The equivalence of the unfolded
program and the original (non annotated) one is proven (Proposition~\ref{lemma:nequality}),
by using  a slightly modified operational
semantics for annotated programs (as defined in Section~\ref{sec:semantics}).
We then provided a condition that could be used to safely replace a
rule with its unfolded version, whilst simultaneously preserving qualified answers, for a
restricted class of rules. Confluence and termination maintenance of
the program modified in the previous way are proven.

There are only few other papers that consider source to source transformation
of CHR programs. \cite{Fru04}, rather than considering a generic transformation
system focuses on the specialization of rules regarding a specific goal, analogously
to what happens in partial evaluation. In \cite{FH03}, CHR rules are transformed
in a relational normal form, over which a source to source transformation is performed.
However, the correctness of such a transformation was not proven. Some form of
transformation for probabilistic CHR is considered in \cite{FDPW02}, while
guard optimization was studied in \cite{SSD05b}.

Both the general and the goal specific approaches are important in order
to define practical transformation systems for CHR. In fact, on the
one hand of course one needs some general unfold rule, on the other
hand, given the difficulties in removing rules from the transformed
program, some goal specific techniques can help to improve the
efficiency of the transformed program for specific classes of
goals. A method for deleting redundant CHR rules is considered
in \cite{AF04}. However it is based on a semantic check and it
is not clear whether it can be transformed into a specific syntactic
program transformation rule.

When considering more generally the field of concurrent logic
languages, we find a few papers which address the issue of program
transformation. Notable examples include  \cite{EGM01} that deals with
the transformation of  concurrent constraint programming (ccp) and
\cite{UF88} that considers Guarded Horn Clauses (GHC).  The
results in these papers are not directly applicable to CHR  because
neither ccp not GHC allow rules with  multiple heads.

The third section of this paper can be considered as a first step
in the direction of defining a transformation system for CHR programs,
based on unfolding. This step could be improved in several directions.
First of all, the unfolding operation could be extended to also take
into consideration the constraints in the propagation part of the head
of a rule instead of only the body ones.
In addition, the condition that we have provided for safely
replacing a rule could be generalized to include more cases.
Also, we could extend to CHR some of the other transformations,
notably folding, which have been defined  in \cite{EGM01} for ccp.
Finally, we would like to investigate from a practical perspective
to what extent program transformation can improve the
performances of the CHR solver. Clearly the application of an
unfolded rule avoids some computational steps assuming of course that
unfolding is done at the time of compilation, even though the
increase in the number of rules could eliminate this improvement
when the original rule cannot be removed.
Here it would probably be important to consider some unfolding
strategy, in order to decide which rules have to be unfolded.

\bibliographystyle{acmtrans}

\end{document}